\documentclass[aps,pre,final,twocolumn,groupedaddress]{revtex4-1}
\usepackage[english]{babel}
\usepackage[utf8]{inputenc} 
\usepackage[T1]{fontenc} 
\usepackage{amsmath,amssymb,amsthm,relsize,bbm}
\usepackage{xspace} 
\usepackage{graphicx}
\usepackage{enumitem}
\usepackage{xcolor} 
\usepackage{soul} 

\newcommand{\functiondef}[5]{\begin{equation} \begin{array}{r@{\,}r@{\;}c@{\;}l}
#1 \colon & #2 & \mathlarger{\rightarrow} & #3 \\ & #4 & \mathlarger{\mapsto} & #5 \end{array}\end{equation}}
\newcommand{\ql}{\textquotedblleft{}\ignorespaces}
\newcommand{\qr}{\textquotedblright{}\xspace}
\newcommand{\qsl}{\textquoteleft{}\ignorespaces}
\newcommand{\qsr}{\textquoteright{}\xspace}

\newcommand{\Eq}{Eq.~}
\newcommand{\Eqs}{Eqs.~}
\newcommand{\eqdef}{\stackrel{\mathrm{def}}{=}}
\newcommand{\di}{\mathrm{d}}
\newcommand{\p}[2][\phantom{.}]{\dfrac{\partial #1}{\partial #2}}
\newcommand{\f}[2][\phantom{.}]{\dfrac{\delta #1}{\delta #2}}

\newcommand{\rr}{\boldsymbol{r}}
\newcommand{\pp}{\boldsymbol{p}}
\newcommand{\qq}{\boldsymbol{q}}

\newcommand{\PPsi}{\boldsymbol{\Psi}}
\newcommand{\bbeta}{\boldsymbol{\beta}}
\newcommand{\cc}{\boldsymbol{c}}
\newcommand{\CC}{\boldsymbol{C}}
\newcommand{\JJ}{\boldsymbol{J}}

\newcommand{\ig}{\rotatebox[origin=c]{180}{$g$}}
\newcommand{\Dirac}{\delta^{(3)}}

\renewcommand{\iiint}{\!\int\!\!}

\DeclareMathOperator{\grad}{grad}
\DeclareMathOperator{\vspan}{span}
\DeclareMathOperator{\Tr}{Tr}

\newtheorem{def.}{Definition}[section]
\newtheorem{th.}[def.]{Theorem}
\newtheorem{cor.}[def.]{Corollary}

\begin{document}

\title{Essential  equivalence of the GENERIC and  Steepest Entropy Ascent  models  of dissipation for   non-equilibrium thermodynamics }

\author{Alberto Montefusco and Francesco Consonni}
\affiliation{
Politecnico di Milano, Via Ponzio 34/3, Milano, Italy}

\author{Gian Paolo Beretta}
\affiliation{
Universit\`a di Brescia, via Branze 38,
Brescia, Italy}

\email{gianpaolo.beretta@unibs.it}
\date{\today}

\begin{abstract}
By reformulating the Steepest-Entropy-Ascent (SEA) dynamical model for non-equilibrium thermodynamics in the mathematical language of Differential Geometry, we  compare it with the primitive  formulation of the GENERIC model and discuss the main technical differences of the two approaches. In both dynamical models the description of dissipation is of the \qsl entropy-gradient\qsr type. SEA focuses only onto the irreversible component of the time evolution, chooses a  sub-Riemannian metric tensor as dissipative structure, and uses the local entropy density field as potential. GENERIC emphasizes the coupling between the reversible and irreversible components of the time evolution, chooses two compatible degenerate structures (Poisson and degenerate co-Riemannian), and uses the global energy and entropy functionals as potentials. As an illustration, we rewrite the known GENERIC formulation of the Boltzmann Equation in terms of the square-root of the distribution function adopted by the SEA formulation.  We then provide a formal proof that in more general frameworks, whenever all degeneracies in the GENERIC framework are related to conservation laws , the SEA and GENERIC models of the irreversible component of the dynamics are essentially interchangeable, provided of course they assume the same kinematics. As part of the discussion, we note that equipping the dissipative structure of GENERIC with the Leibniz identity makes it automatically
SEA on metric leaves.
\end{abstract}

\maketitle

\section{Introduction}\label{Intro}

The basic concepts and applications of equilibrium thermodynamics are among the most  consolidated milestones in physics. On the other hand,  thermodynamic theories capable of describing non-equilibrium states and their time evolution are still at the forefront of research in physics, fostered by a wide variety of applications and new technologies that are in need of such modeling capability.

Various theories and approaches to non-equilibrium dynamics have been put forward since the pioneering work by Onsager in 1931 \cite{lO31}. It is not our purpose here to review such huge scientific literature, nor to even acknowledge the many pioneers of this broad topic. Therefore, the reader interested in such reconstructions should not start from our list of references.

The scope of this paper is to compare two quite general geometrical constructions that have evolved independently, with somewhat different purposes, but that turn out to provide almost equivalent -- or at least very closely related -- dissipative structures that guarantee the compatibility of a non-equilibrium thermodynamics theory with the second law of thermodynamics.

At any level of description, the  geometrization of a theory of non-equilibrium (thermo)dynamics consists in identifying: (a) the state space $\mathcal{M}$ assumed for the physical system under study, (b) the structure of this space, (c) the time evolution equation in terms of this structure, (d) the compatibility of dynamics with the statement of the second law  of thermodynamics \cite{gpB86b}, and (e)  the symmetry group of the theory, i.e., the group preserving the geometrical structure of $\mathcal{M}$ \cite{rM00}.

Not all existing non-equilibrium thermodynamics theories have been clearly geometrized yet in this sense. However, much progress along these lines has been made and constitutes the background of the present work:
\begin{itemize}[leftmargin=*]
\item Classical Mechanics has been formulated in an abstract (general) setting, in the context of Geometric Mechanics (\cite{viA89,MR03}): the natural arenas are symplectic manifolds, and their generalization, i.e., Poisson manifolds.
\item Equilibrium Thermodynamics has been  geometrized in the work by Carathéodory \cite{cC09}, the book by Hermann \cite{rH73}, and -- for example -- the references in \cite{hQ07}.
\item Some formulations of Non-equilibrium Thermodynamics have been reformulated using the important geometric structure of \emph{metriplectic manifolds} (see some history and references in \cite{pjM09} and \cite{dF05}).
\item In its most renowned presentation, metriplectic dynamics has been called \emph{General Equation for the Non-Equilibrium Reversible-Irreversible Coupling} (GENERIC) \cite{GO97}, which represents also a generalization in the context of \emph{contact manifolds}.
\item An apparently less structured approach, \emph{Steepest-Entropy-Ascent} SEA dynamics, was proposed in the simplest quantum thermodynamic landscape \cite{gpB86,gpB87,sGS01,sGS01add,gpB09,CBS15} and in a general probabilistic framework \cite{gpbfrontiers86,ASME86,ASME87Boston,ASME87Rome,gpB08}, and recently adapted and generalized for meso- and macroscopic systems in \cite{gpB14}.
\end{itemize}

In general, a reversible evolution is modeled with an antisymmetric tensor, while the irreversible one with a symmetric object. When the latter is a tensor, clearly it may be represented by a kind of metric tensor. This is why below we shall associate dissipation with a metric tensor.

	Although we shall focus here only on the mathematical formulation, it is worth noting that  thermodynamic theories (equilibrium and non-equilibrium) have often been attached very different physical (and philosophical) interpretations. For example:
	\begin{itemize}[leftmargin=*]
	\item The Keenan school of thermodynamics has advanced the position that thermodynamics is valid at every scale and that entropy is an intrinsic property of matter which, like energy, builds up from the microscopic level. The SEA geometrical approach originated from probing the extreme consequences of this line of reasoning, to seeing how the quantum dynamics should be modified if entropy and dissipation existed even microscopically.
	\item On the other hand, the multiscale dynamics advanced by Grmela sees Thermodynamics as \ql \textsl{a meta-theory addressing relations among dynamical theories formulated on different levels of description}\qr \cite{mG13}; these relations are expressed in the framework of contact structures \cite{mG14}. At every scale there is a GENERIC, and in passing from a more detailed level to a less detailed one through \qsl pattern recognition\qsr (or coarse graining \cite{hcO05})  one sees dissipation, even if microscopic dynamics is reversible.
	\end{itemize}
	Our scope here is not to elaborate on any of these interpretational issues, nor on the operational definitions and empirical meanings of basic concepts such as energy and entropy. We take the view that, regardless of their interpretation, two theories are identical if their mathematical structures are equivalent.

The paper is structured as follows. In Section \ref{MathForm} we reformulate the SEA and the GENERIC approaches so as to emphasize their analogies and differences, which we further discuss in Section \ref{Discussion}. In Section \ref{BE} we implement our formalism to the case of the Boltzmann Equation. In Section \ref{seaVSgeneric} we further elaborate our notation and analysis so as to establish when and  in what sense SEA and  GENERIC can be considered equivalent. In Section \ref{Conclusions} we summarize our conclusions.

\section{Mathematical Formulation}\label{MathForm}
In this section, we present a mathematical and abstract formulation of the SEA and GENERIC principles that allows a clear comparison between the respective underlying assumptions. To help the reader go through the mathematics and grasp the general meaning without getting sidetracked by the details, we will try to guide the reading as much as possible by adding some non-technical comments that, albeit strictly unnecessary from the mathematical point of view, are meant to allow a simpler interpretation of the abstract setup.

We will focus on the dynamics. Regarding the kinematics, for both SEA and  GENERIC we will assume it to be given. In particular, when in Section \ref{seaVSgeneric} we will derive  relations between the structures of dissipative dynamics in SEA and GENERIC models, we will do under the assumption  that the two models adopt the same kinematics. In general, for both SEA and  GENERIC, the kinematics is  chosen so that the state space is  a Banach manifold $\mathcal{M}$, i.e., a manifold which -- locally -- is topologically equivalent to a Banach space.

The time evolution of the state is represented by a curve $\alpha: I \rightarrow \mathcal{M} \left( I \subseteq \mathbb{R} \right)$ on $\mathcal{M}$, and this is an integral curve of a vector field (the velocity vector is equal to the vector field at each point of the curve). The vector field is composed of two distinguished parts: the first one is a \emph{reversible} contribution, $X^H$, which, depending on the framework, represents Hamiltonian dynamics and/or the local effects of transport due to convective and diffusive fluxes between adjacent elements of a continuum; the second one is a \emph{dissipative} contribution, which  models the irreversible aspects of the dynamics (such as the dissipation of mechanical or electrical forms of energy into thermal energy) responsible for the  entropy production (the local entropy production in the case of a continuum).  In symbols, the time evolution $\alpha(t)$ is an integral curve of the sum of the reversible and dissipative vector fields,
\begin{equation}\label{DE}
\dot{\alpha}(t) = X^H_{\alpha(t)} + Y^S_{\alpha(t)} \ .
\end{equation}

Our comparison between the SEA and the GENERIC constructions will  focus the attention on the dissipative part because this has been the  focus of the SEA construction, namely to define the dissipative part $Y^S$ in non-equilibrium frameworks where the reversible or transport part $X^H$ is prescribed by other considerations. Instead, the GENERIC construction provides specifications also for $X^H$.

As we will see below in detail, both the SEA and the GENERIC constructions assume that the dissipative part $Y^S$ of the dynamical equation \eqref{DE} is directly related to the entropy differential. In the SEA construction, it is related to the projection of the gradient of the local entropy functional onto a linear manifold orthogonal to the gradients of the local functionals representing the conserved properties. In the GENERIC construction we consider in this article (where $\mathcal{M}$ is a metriplectic manifold), it is related to a weaker notion of \qsl gradient\qsr.

First of all, also to establish the notation, we recall some basic notions of differential geometry. The most useful definition of (tangent) vector on a manifold passes through the concept of \emph{derivation}. A \emph{tangent vector} $v_p$ to a point $p$ of the manifold $\mathcal{M}$ is a \emph{derivation} on $C^\infty(\mathcal{M})$; that is, a linear map
\functiondef{v_p}{C^\infty(\mathcal{M})}{\mathbb{R}}{ A }{v_p( A ) \ ,}
which takes any smooth functional $A$ on $\mathcal{M}$, gives a real number $v_p(A)$ and satisfies the \emph{Leibniz rule}
\begin{equation}\label{Leibniz1}
v_p(AB) = v_p(A)\, B_p + A_p\, v_p(B)
\end{equation}
for any functionals $A$ and $B$, where $A_p$ and $B_p$ denote their values at $p$. The set of all tangent vectors to $p$ is a vector space, called the \emph{tangent space to $\mathcal{M}$ at $p$} and denoted by $T_p\mathcal{M}$. The disjoint union of all tangent spaces is the \emph{tangent bundle}, denoted by $T\mathcal{M}$. A \emph{vector field} $X$ is a map
\functiondef{X}{\mathcal{M}}{T\mathcal{M}}{p}{X_p \ ,}
with the property that $X_p \in T_p\mathcal{M}$ $\forall p \in \mathcal{M}$: i.e., it assigns a tangent vector $X_p$ in $T_p\mathcal{M}$ to each point $p$ of $\mathcal{M}$.

By referring to the model space of the manifold, one may express a tangent vector by the linear combination
\begin{equation}
v_p = v_p^k \partial_k ,
\end{equation}
where $\{\partial_k\}$ is a basis of derivations for the tangent space, or -- in other words -- the partial derivatives with respect to the coordinates. By the same consideration, one is allowed to see $v_p(A)$ as the \emph{directional derivative} of $A$ at $p$ along the tangent vector $v_p$.

Moreover, one may define the \emph{cotangent space at $p$} ($T^*_p\mathcal{M} \eqdef (T_p\mathcal{M})^*$) as the space of all linear functionals (named \emph{covectors}) on $T_p\mathcal{M}$. The disjoint union of all cotangent spaces is the \emph{cotangent bundle} $T^*\mathcal{M}$. The most important of these linear functionals is the \emph{differential} $\di A$ of a smooth functional  $A$ on $\mathcal{M}$, which computes the directional derivatives of $A$ at every point $p$ of $\mathcal{M}$ along the  tangent vectors to $p$, i.e., at each point $p$  takes any tangent vector $v_p$ as input and yields the directional derivative $v_p(A)$ as output,
\begin{equation}
\di A_p(v_p) = v_p(A)
\end{equation}

The notion of gradient of a smooth functional requires an additional structure on the manifold. Indeed, it can be defined in an invariant way (i.e., independent of the choice of coordinates) only with respect to a non-degenerate bilinear map on the tangent bundle (or some sub-bundle), which essentially equips the manifold with a metric.  The most common case is represented by \emph{(strong) Riemannian manifolds}. These are defined as pairs of a smooth manifold and a  (strongly non-degenerate) Riemannian metric tensor field $g_p$ which, at every point $p$ of $\mathcal{M}$ takes as input two vectors in the tangent space $T_p\mathcal{M}$, and yields
\functiondef{g_p}{T_p\mathcal{M} \times T_p\mathcal{M}}{\mathbb{R}}{(u_p, v_p)}{g_p(u_p, v_p)}
with $g_p(u_p, v_p)>0$ for any nonzero $u_p$ and $v_p$.
The property of strong non-degeneracy implies that the vector bundle (linear) map (at every point $p$) $g_p^\flat : T_p\mathcal{M} \rightarrow T_p^*\mathcal{M}$, defined by
\begin{equation}
[g_p^\flat (u_p)] (v_p) = g_p(u_p, v_p)  \qquad \forall  v_p \in T_p\mathcal{M}\ ,
\end{equation}
which brings a vector $u_p$ into the covector $g_p^\flat (u_p)$ (i.e., into a linear functional on the tangent space at $p$), is an isomorphism (often called \emph{musical isomorphism}). Therefore, the inverse map $g_p^\sharp : T_p^*\mathcal{M} \rightarrow T_p\mathcal{M}$ is defined too.  We may also define the \qsl inverse\qsr metric tensor, or \emph{co-metric tensor} by
\begin{equation}
\ig_p(\omega_p, \eta_p) = g_p \!\left(g^\sharp_p (\omega_p), g^\sharp_p (\eta_p) \right)\ .
\end{equation}

 In coordinates and finite dimensions, a physicist would talk about \emph{lowering} and \emph{raising the indexes}:
\begin{equation}
\big( g_p^\flat \big)_{ij} v_p^j = v_{p,i} , \qquad \big( g_p^\sharp \big)^{ij} v_{p,j} = v_p^i ,
\end{equation}
where $v^j$ and $v_j$ are respectively the components of a vector and a covector with respect to some chosen basis for the tangent space $T_p\mathcal{M}$ and the cotangent space $T^*_p\mathcal{M}$; $\left( g_p^\flat \right)_{ij} = g_{p, ij}$ and $\left( g_p^\sharp \right)^{ij} = g_p^{ij}$ are the matrix representations of the maps $g_p^\flat$ and $g_p^\sharp$ with respect to the same bases, and of course $\left[g_p^{ij}\right] = \left[g_{p,ij}\right]^{-1}$.

One then defines the \emph{gradient} at $p$ of a smooth functional $A$ to be the only vector at $p$ satisfying
\begin{equation}\label{gradient}
\di A_p(v_p) = g_p\left( \grad_g A|_p, v_p \right) \quad \forall  v_p \in T_p\mathcal{M} \mbox{ and } p \in \mathcal{M} \ .
\end{equation}
Uniqueness is guaranteed by the non-degeneracy of the metric field. This may also be restated more explicitly as
\begin{equation}\label{gradientdiesis}
\grad_g A|_p = g^\sharp_p (\di A_p) \ .
\end{equation}

When $\mathcal{M}$ is a vector space $V$, the tangent space to each point $p$ may be identified with the vector space itself (we write $T_pV \cong V \ \forall p$). Moreover, if the vector space is a Hilbert space $\mathcal{H}$, it is equipped with an inner product, i.e., a non-degenerate bilinear map
\functiondef{\langle , \rangle}{\mathcal{H} \times \mathcal{H}}{\mathbb{R}}{(p, q)}{\langle p , q \rangle \ .}
If we take the manifold viewpoint, this may also be seen as
\functiondef{\langle , \rangle}{T_p\mathcal{H} \times T_p\mathcal{H}}{\mathbb{R}}{(u_p, v_p)}{\langle u_p , v_p \rangle \ ,}
since $T_p\mathcal{H} \cong \mathcal{H} \ \forall p$. Then the particular  \emph{gradient} of a functional $A$ at point $q$, called \emph{variational derivative} of $A$, is  defined implicitly by
\begin{equation}\label{gradientI}
\di A_{q}(v_q) = \left\langle\! \left.\frac{\delta A}{\delta p}\right|_{q}, v_q \right\rangle \quad \forall  q \in \mathcal{H} \ \text{and} \ v_q \in T_q\mathcal{H} \ (\cong \mathcal{H}) \ .
\end{equation}
Given the inner product, we may denote by $R_p$ the Riesz isomorphism $R_p \colon T^*_p\mathcal{H} \to T_p\mathcal{H}$ such that
\begin{equation}
\left\langle R_p(\omega_p), v_p \right\rangle = \omega_p(v_p) \quad \forall v_p \in T_p\mathcal{H}, \ \forall \omega_p \in T^*_p\mathcal{H} \ ,
\label{isomorphism}
\end{equation} and hence we may alternatively define the variational derivative explicitly by
\begin{equation}\label{variationalderivative}
\left.\f[A]{p}\right|_{q} \eqdef R_{q} (\di A_{q}) \ .
\end{equation}

When both structures (an inner product and a metric) are present on a Hilbert space $\mathcal{H}$, we have
\begin{equation}
\grad_g A|_p =g_p^\sharp \!\left(\! R_p^{-1}\left(\! \left.\frac{\delta A}{\delta p}\right|_p \right)\!\right)= \hat{G}_p^{-1} \!\left(\! \left.\frac{\delta A}{\delta p}\right|_p \right)
\label{varder2grda}
\end{equation}
where $\hat G_p^{-1}$ denotes the inverse of the positive definite and symmetric linear operator $\hat G : T_p\mathcal{M} \rightarrow T_p\mathcal{M}$ defined by
\begin{equation}
\hat{G}_p (v_p) \eqdef R_p \!\left( g^\flat_p (v_p) \right) \qquad \forall v_p \in T_p\mathcal{H} \ .
\end{equation}

\subsection{Steepest Entropy Ascent}\label{SEAgeneral}

The Steepest Entropy Ascent (SEA) principle to model the non-equilibrium dynamics of a thermodynamic system was originally proposed as part of an attempt to design a thermodynamically consistent dynamical law for a unified quantum theory of mechanics and thermodynamics obtained by embedding the second law directly into  the set of fundamental  postulates \cite{gpB84,gpB85,gpB85b,gpB85c,jM85,gpB86,sGS01,sGS01add,gpB09}. Subsequently it was extended as a generic modeling tool for probabilistic, constrained maximum entropy landscapes \cite{gpB87,gpB06,gpB08}. Since in these landscapes the state representative is a probability distribution or its quantum equivalent, the density operator, it seemed natural to define gradients with respect to the Fisher-Rao metric which is   known to exhibit the required invariance features in the absence of additional physical constraints. However, the Fisher metric is not general enough to reproduce the dynamics in  other non-equilibrium thermodynamics frameworks. Therefore, an extension of the SEA formulation that adopts a generic metric has been provided in \cite{gpB14}. In this section we present the structure of such more general formulation.  Actually, as part of our effort to cast the construction in  the language of differential geometry, we first  present a further generalization whereby the state space is not   required to be a vector space but may be any (Banach) manifold, then we present the original formulation where the state manifold is equipped with an inner product so that variational derivatives are defined.

The idea behind the SEA construction is to \qsl geometrize\qsr the thermodynamic state space and assume that the part of the local evolution  equation that is responsible for irreversible dynamics is in the \qsl direction\qsr of maximal entropy production compatible with the local conservation constraints. The formulation of the SEA principle may be expressed as follows:
\emph{the
  time evolution of the local state is the result of a balance between the effects of transport or Hamiltonian dynamics and the spontaneous and irreversible tendency to advance the local state representative in the direction of maximal  entropy production per unit of distance  traveled in state space compatible with the conservation constraints} \cite{gpB14}.

  The mathematical implementation consists in assuming that the dissipative part of the dynamics pushes the states in the direction of the  gradient of the restriction of the entropy functional onto the submanifold with constant values of the conserved quantities. Therefore, the dissipative vector field is assumed to point in the direction of maximal directional derivative of the entropy compatible with the conservation constraints.

\subsubsection{Generalized abstract formulation}\label{SEAgeneralized}

Each thermodynamically consistent non-equilibrium theory assumes a level of description for a given physical system (possibly modeled as a continuum) which mathematically amounts to assuming:
\begin{itemize}[leftmargin=*]
\item a (possibly infinite-dimensional) smooth real Banach manifold $\mathcal{M}$ whose points represent the possible  states of the system  or, for nonequilibrium states of a continuum, the possible local states at position $\qq$;
\item a set of functionals $c^i \colon \mathcal{M} \rightarrow \mathbb{R}$, which represent the conserved properties of the system or, for a continuum, the local densities of the conserved properties;  we denote the submanifold $\{p \in \mathcal{M} \colon c^i(p) = \text{const}_i \ \forall i\}$ with $\mathcal{M}_{\cc}$;
\item another functional $s \colon \mathcal{M} \rightarrow \mathbb{R}$, which represents the thermodynamic entropy \cite{foot1} of the system or, for a continuum, the local entropy density;
\item for each submanifold $\mathcal{M}_{\cc}$, a (strongly non-degenerate) metric tensor field $g$ which, at every point $p$ of the submanifold $\mathcal{M}_{\cc(p)}$, takes as input two vectors in the tangent space $T_p\mathcal{M}_{\cc(p)}$, and yields
\functiondef{g_p}{T_p\mathcal{M}_{\cc(p)} \times T_p\mathcal{M}_{\cc(p)}}{\mathbb{R}}{(u_p, v_p)}{g_p(u_p, v_p)}
with $g_p(u_p, v_p)>0$ for any nonzero $u_p$ and $v_p$, and is such that the map $p \mapsto g_p$ defines a smooth ($C^\infty$) map on  $\mathcal{M}$ .  The condition that defines the space $T_p\mathcal{M}_{\cc(p)}$  tangent to the constrained submanifold $\mathcal{M}_{\cc(p)}$ is
\begin{equation}\label{defTpMc}
T_p\mathcal{M}_{\cc(p)}=\{v_p^{\cc}\in T_p\mathcal{M} : \di c^i_p(v_p^{\cc}) = 0\  \forall i \} \ .
\end{equation}
\end{itemize}
This situation essentially defines a \emph{sub-Riemannian} structure on $\mathcal{M}$. We shall come back to this point in Subsection \ref{geometry}.

The above assumptions allow one to define the gradient  with  respect to the given metric field $g$ of the smooth functional  $s^{\cc} : \mathcal{M}_{\cc}  \rightarrow \mathbb{R}$ defined by the restriction of the functional $s \colon \mathcal{M}  \rightarrow \mathbb{R}$ on the submanifold  $\mathcal{M}_{\cc}$ of $\mathcal{M}$ where the conserved functionals $c^i$ are constrained to fixed constant values $\cc$, i.e.,  such that
\begin{equation}
v_p^{\cc} \mapsto \di s^{\cc}(v_p^{\cc})=\di s(v_p^{\cc}) \ .
\end{equation}
  We can do it either through the following implicit expression (\Eq\eqref{gradient}): $\grad_g^{\cc} s^{\cc}|_p$ is the unique vector in $T_p\mathcal{M}_{\cc(p)}$ such that
\begin{equation}\label{gradientdef}
\di s_p^{\cc}(v_p^{\cc}) = g_p(\grad_g^{\cc} s^{\cc}|_p, v_p^{\cc}) \qquad \forall  v_p^{\cc} \in T_p\mathcal{M}_{\cc(p)}\ ;
\end{equation}
or, explicitly,
\begin{equation}
\left. \grad_g^{\cc} s^{\cc} \right\rvert_p = g_p^\sharp \left(\di s_p^{\cc}\right) .
\label{gradient_expl}
\end{equation}

With reference to the time evolution equation \eqref{DE}, the SEA construction  focuses only on the dissipative vector field $Y^S$ because its objective is to construct a dynamics in which the entropy functional $s$ is an $S$-function in the sense defined in Ref.\ \cite{gpB86b}  so that the maximal entropy states (or, for a continuum, the locally maximal entropy density states) represent the only stable (local) equilibrium states of the system, consistently with the Hatsopoulos-Keenan statement of the Second Law \cite{HK65,GB05}. Instead, the non-dissipative vector field $X^H$ in \Eq\eqref{DE}, which represents the reversible components of the system dynamics or, for a continuum, the local net effects of transport of properties between adjacent elements of the continuum, are assumed to be given features of the level of description and coarse graining of the modeling framework in which the SEA construction is to be implemented.

The SEA dynamics is obtained by assuming  the dissipative vector field $Y^S$ as follows
\begin{equation}
Y^S_{\alpha(t)} =\dfrac{1}{\tau}  \grad_g^{\cc(\alpha(t))} s^{\cc(\alpha(t))}|_{\alpha(t)} =\dfrac{1}{\tau} g^\sharp_{\alpha(t)}\!\!\left( \di s_{\alpha(t)}^{\cc(\alpha(t))} \right) ,
\label{SEA}
\end{equation}
where $\tau$ is a positive dimensionality constant. Since by definition (\ref{gradientdef}) the vector $\grad_{\cc} s^{\cc}|_p $ is in $T_p\mathcal{M}_{\cc(p)}$ and therefore, recalling \Eq\eqref{defTpMc}, is in the kernel of every $\di c^i$, \Eq\eqref{SEA} satisfies automatically the following conservation constraints
\begin{equation}\label{conservation}
\di c^i_{\alpha(t)}\left(\dot{\alpha}(t)-X^H_{\alpha(t)}\right) = \di c^i_{\alpha(t)} \left( Y^S_{\alpha(t)} \right)  = 0 \ .
\end{equation}

As discussed in \cite{gpB14} \Eq\eqref{SEA} can also be viewed as the solution of a maximal entropy production variational problem that in the notation just developed can be expressed as follows
\begin{equation}\label{SEAvariational}
\max_{Y^S}|_p\  \di s_p^{\cc(p)}(Y^S) \mbox{ subject to } g_p(Y^S,Y^S)=\mbox{const}
\end{equation}

The rate of entropy production can be expressed  in the following equivalent forms
\begin{align}
& \di s_{\alpha(t)} \left( \dot{\alpha}(t)-X^H_{\alpha(t)} \right)=\di s_{\alpha(t)}\left( Y^S_{\alpha(t)} \right) \nonumber \\
&=\di s_{\alpha(t)}^{\cc(\alpha(t))}\left( Y^S_{\alpha(t)} \right) \nonumber \\
&= g_{\alpha(t)}\!\! \left(\grad_g^{\cc} s^{\cc(\alpha(t))}|_{\alpha(t)} , Y^S_{\alpha(t)} \right) \nonumber \\
&= \dfrac{1}{\tau}\, g_{\alpha(t)}\! \left( \grad_g^{\cc} s^{\cc(\alpha(t))}|_{\alpha(t)} , \grad_g^{\cc} s^{\cc(\alpha(t))}|_{\alpha(t)} \right) \nonumber \\
&=\tau\,g_{\alpha(t)}\!\!\left( Y^S_{\alpha(t)},Y^S_{\alpha(t)} \right)  \geq 0 \ ,
\end{align}
where the second equality follows from \Eq\eqref{conservation}, the third from \Eq\eqref{diffFc}, the fourth from \Eq\eqref{gradientdef},  the inequality from the positivity of the metric tensor, and the strict equality holds if and only if $Y^S_{\alpha(t)}=0$. The thermodynamic principle of impossibility of a negative entropy production is thus automatically satisfied.

This thermodynamic consistency feature is very often identified with  the  Second Law. However, the Second Law requires additionally that among the \emph{non-dissipative states}, i.e., the states $p_\text{nd}$ for which $Y^S_{p_\text{nd}}=0$, only those that have maximal entropy for the given values of the conserved properties should be stable equilibrium states with respect to a purely dissipative dynamics, i.e., \Eq\eqref{DE} in which we set $X^H=0$. In other words, we must also prove that all other non-dissipative states, which become equilibrium states when we set  $X^H=0$, should be unstable. This non-trivial additional requirement imposes an additional condition onto the properties that must be satisfied by the entropy functional. It is often stated that entropy provides a Lyapunov criterion for the stability of the thermodynamic (local) equilibrium states. However, as discussed in \cite{gpB86b}, the rigorous justification of this statement is not trivial and requires that the entropy functional satisfies a weaker criterion than that of being a Lyapunov functional which in \cite{gpB86b} is defined precisely and named S-functional. For the quantum framework, the proof of the conjecture advanced in \cite{gpB86b} that the von Neumann entropy functional $-k_\text{B}\Tr\rho\ln\rho$ is indeed an S-functional was later found in \cite{HOT81}.

\subsubsection{Original inner product formulation}\label{SEAoriginal}

The original formulations of the SEA constructions in Refs. \cite{gpB14,sGS01,sGS01add,gpB09,gpB87,gpB06,gpB08} assume that the state manifold  $\mathcal{M}$ is equipped with an inner product $\langle \cdot , \cdot \rangle$. As a consequence,
\begin{itemize}[leftmargin=*]
\item variational derivatives are defined according to \Eq\eqref{gradientI};
\item for shorthand, we denote the variational derivatives of the conserved functionals $c^i \colon \mathcal{M} \rightarrow \mathbb{R}$ with the symbols $\Psi^i$, i.e., \begin{equation}\Psi^i_p\eqdef\left.\frac{\delta c^i}{\delta p}\right|_p\qquad \forall p\in \mathcal{M} \ ;\end{equation}
with the further shorthand symbols $\cc$ and $\PPsi$ to denote the arrays $\cc=\{c^1,c^2,\dots\}$ and  $\PPsi=\{\Psi^1,\Psi^2,\dots\}$, and $\vspan(\PPsi)$ to denote the linear span of the $\Psi^i$'s;
\item again for shorthand, we denote the  variational derivative of the functional $s \colon \mathcal{M} \rightarrow \mathbb{R}$ with the symbol $\Phi$  i.e., \begin{equation}\Phi_p\eqdef\left.\frac{\delta s}{\delta p}\right|_p\qquad \forall p\in \mathcal{M} \ .\end{equation}
\end{itemize}
The gradient  with  respect to the given metric field $g$ of the restriction of the entropy functional  $s^{\cc} \colon \mathcal{M}_{\cc} \to \mathbb{R}$  on the submanifold  $\mathcal{M}_{\cc}$ with constant values of the conserved functionals $c^i$, using  \Eqs\eqref{gradientdiesis} and \eqref{varder2grda}  can be written explicitly as
\begin{equation}
\left. \grad_g^{\cc} s^{\cc} \right\rvert_p = g_p^\sharp \left(\di s_p^{\cc}\right) = \hat G_p^{-1} (\Phi_p^{\cc}) \ .
\end{equation}

Moreover, the tangent space $T_p\mathcal{M}$ at any $p$ on $\mathcal{M}$ is viewed as the orthogonal composition $T_p\mathcal{M}=T_p\mathcal{M}_{\cc(p)}\oplus\vspan(\PPsi_p)$  so that any tangent vector can be decomposed as
\begin{equation}\label{decompvp}v_p=v_p^{\cc}+v_p^{\perp\cc}\end{equation}
with $v_p^{\cc}$ the component tangent to the submanifold  defined by the values at $p$ of the conservation constraints, and $v_p^{\perp\cc}$ the component orthogonal to such submanifold. Indeed, from \Eqs\eqref{defTpMc} and \eqref{gradientI}  written for $A$ being one of the conservation constraints, the condition that defines the space $T_p\mathcal{M}_{\cc(p)}$ tangent to the constrained submanifold (metric leaf) $\mathcal{M}_{\cc(p)}$ becomes
\begin{equation}\label{defTpMc2}
T_p\mathcal{M}_{\cc(p)}=\{v_p^{\cc}\in T_p\mathcal{M} : \di c^i_p(v_p^{\cc}) = \langle \Psi^i|_p, v_p^{\cc} \rangle = 0\  \forall i \} \ ,
\end{equation}
which shows clearly that $T_p\mathcal{M}_{\cc}$ is the orthogonal complement of the linear span of the $\Psi^i$'s.

Along with decomposition (\ref{decompvp}), also the differential  $\di F$ of a smooth  functional  $F$ on $\mathcal{M}$ is naturally decomposed as
\begin{equation}
\di F = \di F^{\cc} + \di F^{\perp\cc} \ ,
\end{equation}
where $\di F^{\cc}$ computes the directional derivative along the component of the tangent vector that lies in $T_p\mathcal{M}_{\cc(p)}$ and $\di F^{\perp\cc}$ along the orthogonal component that lies in $\vspan(\PPsi_p)$, i.e.,
\begin{equation}\label{defconstraineddifferential}
\di F_p^{\cc}(v_p) = \di F(v_p^{\cc}) \mbox{  and  }   \di F_p^{\perp\cc}(v_p)=\di F_p(v_p^{\perp\cc}) \ .
\end{equation}
In particular, when $F=c^i$, definitions (\ref{defTpMc}) and (\ref{defconstraineddifferential}) imply
the identities $\di c^{i,\cc} =0$ and $\di c^{i,\perp\cc}=\di c^i$, i.e.,
\begin{equation}\label{defconstraineddifferentialc}
\di c_p^{i,\cc}(v_p) = 0 \mbox{  and  }   \di c_p^{i,\perp\cc}(v_p)=\di c^i_p(v_p^{\perp\cc})=\di c^i_p(v_p) \ .
\end{equation}

\begin{figure}[h]
\centering
\includegraphics[width=1.0\linewidth]{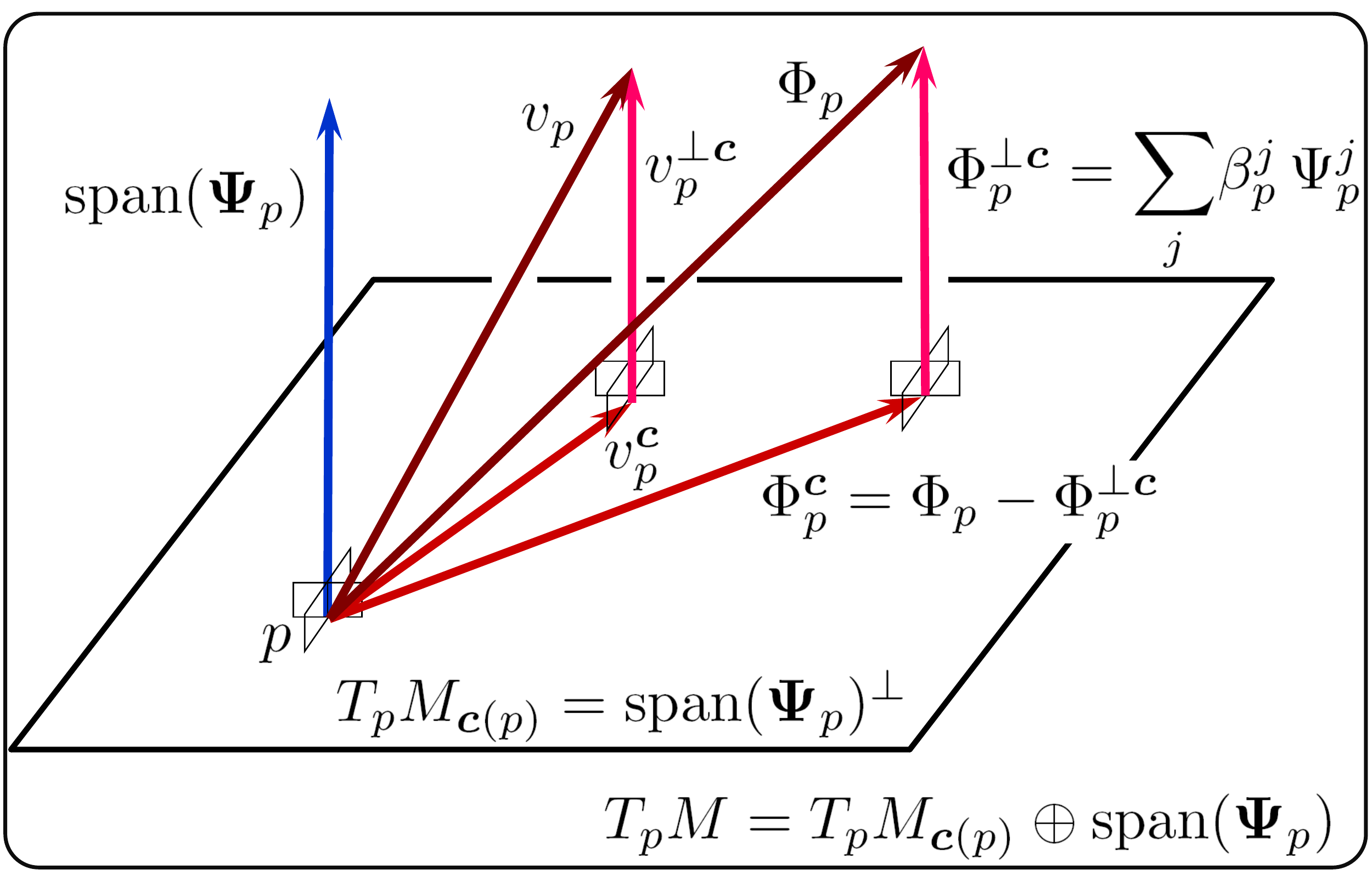}
\caption{Pictorial representation of the orthogonal decompositions $v_p=v_p^{\cc}+v_p^{\perp\cc}$ of a tangent vector  and  $\Phi_p=\Phi_p^{\cc}+\Phi_p^{\perp\cc}$ of the variational derivative  of a smooth functional  $s $ on $\mathcal{M}$ with respect to the  decomposition $T_p\mathcal{M}_{\cc(p)}\oplus {\rm span}(\PPsi_p)$ of the tangent space $T_p\mathcal{M}$ at point $p$ on the state manifold $\mathcal{M}$ where $\mathcal{M}_{\cc(p)}$ is the submanifold with constant values $\cc(p)$ of a set of smooth functionals  $\cc $ on $\mathcal{M}$ with variational derivatives $\PPsi_p$.}
\label{SEA picture}
\end{figure}

Similarly, the decomposition (\ref{decompvp}) for the vector $\Phi_p$ is
\begin{equation}\label{deompchi}
\Phi_p=\Phi_p^{\cc}+\Phi_p^{\perp\cc} \ .
\end{equation}
Since $\Phi_p^{\perp\cc}$ belongs to $\vspan(\PPsi_p)$, there is a set of scalars $\beta^j_p$ such that
\begin{equation}\label{Chiperp}
\Phi_p^{\perp\cc}=\hat{P}_{\vspan(\PPsi_p)} (\Phi_p)= \sum_j\beta^j_p\, \Psi^j \ .
\end{equation}
Recalling \Eq (\ref{gradientI}), we readily see that
\begin{equation}\label{diFperp}
\di s_p^{\perp\cc}(v_p)=\langle \Phi_p^{\perp\cc},v_p\rangle =\sum_j\beta^j_p\, \langle\Psi^j ,v_p\rangle=\sum_j\beta^j_p\, \di c_p^j(v_p) \ ,
\end{equation}
and, therefore, the constrained differential of $s$ and the constrained variational derivatives are, respectively,
\begin{align}\label{diffFc}
\di s_p^{\cc}&= \di s_p -\sum_j\beta_p^j\, \di c_p^j \ ,\\
\label{Xcc}
\Phi_p^{\cc} &= \Phi_p - {\displaystyle \sum_j}\beta_p^j\, \Psi_p^j
\end{align}
where the scalars $\beta^j_p$ are determined uniquely by the solution of the system of equations that obtains by applying \Eq (\ref{diffFc}) to $v_p=\Psi_p^i$ for every $i$, and noting that $\di s_p^{\cc}(\Psi_p^i)=0$ (because clearly $\Psi_p^{i,\cc}=0$), i.e.,
\begin{equation}\label{systemforbetad}
0= \di s_p(\Psi_p^i) -\sum_j\beta_p^j\, \di c_p^j(\Psi_p^i)
\end{equation}
or, equivalently, using again \Eq (\ref{gradientI}),
\begin{equation}\label{systemforbeta}
0= \langle \Phi_p,\Psi_p^i\rangle -\sum_j\beta_p^j \,\langle \Psi_p^j,\Psi_p^i\rangle
\end{equation}
Figure \ref{SEA picture} represents schematically the construction of the constrained variational derivative $\Phi_{\cc}$.

The system of equations \eqref{systemforbeta} defining the multipliers $\beta^j$ can be easily solved, for example using Cramer's rule. Then, following \cite{gpbfrontiers86,ASME86,ASME87Boston,ASME87Rome,gpB08,gpB14}  the $\beta^j$'s can be written explicitly as ratios of determinants so that substitution into \Eq\eqref{diffFc} yields the following explicit expression for the constrained differential
\begin{equation}\label{SEAgram} \di s_p^{\cc}=
\frac{\left|
\begin{array}{cccc} \di s_p & \di \tilde c_p^1 & \cdots& \di \tilde c_p^n \\ \\
\di s_p(\tilde\Psi^1_p) & \di \tilde c_p^1(\tilde\Psi_p^1)  & \cdots & \di \tilde c^1_p(\tilde\Psi_p^n) \\ \\
\vdots & \vdots  & \ddots & \vdots \\ \\
\di s_p(\tilde\Psi_p^n) & \di \tilde c_p^n(\tilde\Psi_p^1)  & \cdots & \di \tilde c^n_p(\tilde\Psi^n_p) \end{array} \right|}{\left|
\begin{array}{ccc}
 \di \tilde c_p^1(\tilde\Psi_p^1)  & \cdots & \di \tilde c_p^1(\tilde\Psi_p^n) \\ \\
 \vdots  & \ddots & \vdots \\ \\
 \di \tilde c_p^n(\tilde\Psi_p^1)  & \cdots & \di \tilde c_p^n(\tilde\Psi_p^n)
 \end{array} \right|}
\end{equation}
and the constrained variational derivative
\begin{equation}\label{dFcgram} \Phi_p^{\cc}=
\frac{\left|
\begin{array}{cccc} \Phi_p & \tilde \Psi_p^1 & \cdots&  \tilde \Psi_p^n \\ \\
\langle \Phi_p,\tilde\Psi_p^1\rangle & \langle \tilde\Psi_p^1,\tilde\Psi_p^1\rangle  & \cdots &  \langle \tilde\Psi_p^1,\tilde\Psi_p^n\rangle \\ \\
\vdots & \vdots  & \ddots & \vdots \\ \\
\langle \Phi_p,\tilde\Psi_p^n\rangle &\langle \tilde\Psi_p^n,\tilde\Psi_p^1\rangle  & \cdots &  \langle \tilde\Psi_p^n,\tilde\Psi_p^n\rangle \end{array} \right|}{\left|
\begin{array}{ccc}
\langle \tilde\Psi_p^1,\tilde\Psi_p^1\rangle  & \cdots &  \langle \tilde\Psi_p^1,\tilde\Psi_p^n\rangle  \\ \\
 \vdots  & \ddots & \vdots \\ \\
\langle \tilde\Psi_p^n,\tilde\Psi_p^1\rangle  & \cdots &  \langle \tilde\Psi_p^n,\tilde\Psi_p^n\rangle
 \end{array} \right|} \ ,
\end{equation}
where we denote by $\tilde c^1,\dots,\tilde c^n$ a subset of the $c^i$'s such that the variational derivatives $\tilde\Psi_p^1,\dots,\tilde\Psi_p^n$  are linearly independent and form a basis for $\vspan(\PPsi_p)$. By virtue of this choice, the determinant at the denominator is a positive definite Gram determinant. Thus, alternatively, we can write
\begin{equation}\label{betaDirect}
\beta^j_p = {\displaystyle \sum_{i=1}^n} [\langle\tilde\PPsi_p,\tilde\PPsi_p\rangle^{-1}]_{ji} \langle \Phi_p,\tilde\Psi_p^i\rangle \ ,
\end{equation}
where, of course, $\langle\tilde\PPsi_p,\tilde\PPsi_p\rangle^{-1} $ denotes the inverse of matrix $\langle\tilde\PPsi_p,\tilde\PPsi_p\rangle$, and $\langle\tilde\PPsi_p,\tilde\PPsi_p\rangle$ is a shorthand to indicate the matrix $[\langle\tilde\Psi^i_p,\tilde\Psi^j_p\rangle]$.

We may also easily construct another set $\overline{c}^1,\dots,\overline{c}^n$ such that the variational derivatives $\overline{\Psi}_p^1,\dots,\overline{\Psi}_p^n$ form an orthonormal basis for  $\vspan(\PPsi_p)$. In such case, we can write
\begin{equation}\label{dFcgramorthonormal}
\di s_p^{\cc} = \di s_p - {\displaystyle \sum_{i=1}^n}\ \langle \Phi_p,\overline{\Psi}_p^i\rangle\, \di \overline{c}_p^i
\end{equation}
and, for the constrained variational derivative of $s$,
\begin{equation}\label{dFcgramorthonormal}
\Phi_p^{\cc} = \Phi_p - {\displaystyle \sum_{i=1}^n}\ \langle \Phi_p,\overline{\Psi}_p^i\rangle\, \overline{\Psi}_p^i
= \Phi_p - \hat{P}_{\vspan(\PPsi_p)} (\Phi_p)
\end{equation}
where
 \begin{align}
 \hat{P}_{\vspan(\PPsi_p)}(\cdot)&=\sum_{i=1}^n\ \langle \cdot,\overline{\Psi}_p^i\rangle\, \overline{\Psi}_p^i \nonumber \\
 &={\displaystyle \sum_{i=1}^n \sum_{j=1}^n} [\langle\tilde\PPsi_p,\tilde\PPsi_p\rangle^{-1}]_{ji} \langle \cdot,\tilde\Psi_p^i\rangle\, \tilde\Psi_p^j
 \end{align}
  is the operator on $T_p\mathcal{M}$ which projects onto $\vspan(\PPsi_p)$.

The SEA  dissipative vector field $Y^S$ becomes
\begin{align}
Y^S_{\alpha(t)} &=\dfrac{1}{\tau}   \grad_g^{\cc} s^{\cc(\alpha(t))}|_{\alpha(t)} =\dfrac{1}{\tau} g^\sharp_{\alpha(t)}\!\!\left( \di s_{\alpha(t)}^{\cc(\alpha(t))} \right) \nonumber  \\ &=
\dfrac{1}{\tau} \hat G^{-1}_{\alpha(t)}\!\! \left( \Phi^{\cc(\alpha(t))}_{\alpha(t)} \right) ,
\end{align}
where $\tau$ is a positive dimensionality constant. The  conservation constraints are
\begin{multline}
\di c^i_{\alpha(t)}\left(\dot{\alpha}(t)-X^H_{\alpha(t)}\right) = \di c^i_{\alpha(t)} \left( Y^S_{\alpha(t)} \right) \\ =  \langle \Psi^i_{\alpha(t)} , Y^S_{\alpha(t)} \rangle = 0 \ .
\label{conservation2}
\end{multline}

The various equivalent expressions for  $ds^{\cc}$ given above  show  explicitly the SEA construction originally introduced in \cite{gpB86,gpbfrontiers86,gpB09} in the quantum thermodynamics framework, whereby the dissipative vector field $Y^S_{\alpha(t)}$ is in the direction of the projection of $\Phi_{\alpha(t)}$ onto the subspace of $T_{\alpha(t)} \mathcal{M}$ orthogonal to all the $\Psi^i|_{\alpha(t)}$'s.

To the list of expressions of the rate of entropy production we can add
\begin{align}
& \di s_{\alpha(t)} \left( \dot{\alpha}(t)-X^H_{\alpha(t)} \right)=\di s_{\alpha(t)}\left( Y^S_{\alpha(t)} \right) \nonumber \\
&=\di s_{\alpha(t)}\left( Y^S_{\alpha(t)} \right)-  \sum\nolimits_{j} \beta^j \,\di  c^j_{\alpha(t)}\left( Y^S_{\alpha(t)} \right) \nonumber \\
&=\tau\,g_{\alpha(t)}\!\!\left( Y^S_{\alpha(t)},Y^S_{\alpha(t)} \right) =
\tau\,\left\langle \hat G_{\alpha(t)}\left(Y^S_{\alpha(t)}\right),Y^S_{\alpha(t)} \right\rangle  \geq 0 \ .
\end{align}

\subsection{GENERIC}\label{GenericGeneral}
In this section we present the simplest form of the GENERIC construction (see \cite{GO97}), the one with the description of entropy production that best resembles the SEA principle.

\subsubsection{Abstract formulation}

We denote by $\mathcal{M}$ the manifold of all possible states $\gamma$ ($\gamma \in \mathcal{M}$) and build the following structure.
\begin{itemize}[leftmargin=*]
\item There exist two potentials, a smooth \emph{Hamiltonian functional} $H \colon \mathcal{M} \rightarrow \mathbb{R}$ and a smooth \emph{entropy functional} $S \colon \mathcal{M} \rightarrow \mathbb{R}$, representing -- for the chosen level of description -- the overall energy and  thermodynamic entropy of the system, respectively.
\item $\mathcal{M}$ is a (possibly infinite-dimensional) Banach \emph{(co)metriplectic} manifold, i.e., a manifold carrying two compatible structures as follows.
\item A Poisson structure describing the reversible part of the dynamics. This is known from Geometric Mechanics (see, for example, \cite{viA89,MR03}), and consists of a skew-symmetric contravariant 2-tensor field, called the \emph{Poisson tensor field}, which -- at every point $p$ of the manifold $\mathcal{M}$ -- takes two covectors at $p$ as inputs, yields
\functiondef{P_p}{T^*_p\mathcal{M} \times T^*_p\mathcal{M}}{\mathbb{R}}{(\omega_p, \eta_p)}{P_p(\omega_p, \eta_p) \ ,}
and is such that the map $p \mapsto P_p$ defines a smooth ($C^\infty$) map on $\mathcal{M}$.
To this tensor we associate a \emph{Poisson bracket} $\{\cdot,\cdot\} \colon C^\infty(\mathcal{M}) \times C^\infty(\mathcal{M}) \to C^\infty(\mathcal{M})$ by the assignment:
\begin{equation}
\{A, B\}_p \eqdef P_p(\di A_p, \di B_p) \ .
\end{equation}
The Poisson bracket must satisfy the Jacobi identity
\begin{equation}\label{JacobyPoisson}
\{A, \{B, C\}\} + \{B, \{C, A\}\} + \{C, \{A, B\}\} = 0 \ ,
\end{equation}
which represents a further constraint on the Poisson tensor field $P_p$. The Poisson tensor also yields the vector bundle (linear) map $P_p^\sharp : T_p^*\mathcal{M} \rightarrow T_p^{**}\mathcal{M}$, called \emph{Poisson operator},  and often assumed to satisfy the condition $P_p^\sharp(T_p^*\mathcal{M}) \subseteq T_p^{**}\mathcal{M} \subseteq T_p\mathcal{M}$, which is needed \cite{OR03} to guarantee that $P_p^\sharp(\di H_p)$ is a vector field. This condition is automatically satisfied whenever the manifold is modeled on a reflexive Banach space or, as a particular case, on a Hilbert space, such as in the Boltzmann Equation framework that we discuss as an example in Section \ref{BE}. Since $P_p$ is assumed to be possibly degenerate, $P_p^\sharp$ is in general non-invertible (it is not a vector-space isomorphism, but only a homomorphism).

It is noteworthy that  condition (\ref{JacobyPoisson}) is imposed so as to implement time-translation invariance of the reversible part of the dynamics (integrability condition). We shall see below that GENERIC does not impose an analogous integrability condition on the irreversible part of the dynamics. We further discuss this point in Section \ref{geometry}.
\item A degenerate co-Riemannian structure (i.e., we have a degenerate \emph{co-}metric instead of a non-degenerate metric) describing the irreversible dynamics \cite{foot2}. This consists of a symmetric and non-negative definite contravariant 2-tensor field, called the \emph{friction tensor field}, which -- at every point $p$ of the manifold $\mathcal{M}$ --	takes two covectors at $p$ as inputs, yields
\functiondef{D_p}{T^*_p\mathcal{M} \times T^*_p\mathcal{M}}{\mathbb{R}}{(\omega_p, \eta_p)}{D_p(\omega_p, \eta_p) \ ,}
and is such that the map $p \mapsto D_p$ defines a smooth ($C^\infty$) map on $\mathcal{M}$.

This tensor equips the set of smooth functionals on $\mathcal{M}$ with the \emph{dissipative bracket} $[\cdot,\cdot]  \colon C^\infty(\mathcal{M}) \times C^\infty(\mathcal{M}) \to C^\infty(\mathcal{M})$ by the assignment:
\begin{equation}
[A, B]_p \eqdef D_p(\di A_p, \di B_p) \ .
\end{equation}
The friction tensor also yields the vector bundle map $D_p^\sharp : T_p^*\mathcal{M} \rightarrow T_p\mathcal{M}$, called \emph{friction operator}, also often assumed to satisfy the condition $D^\sharp(T^*\mathcal{M}) \subseteq T^{**}\mathcal{M} \subseteq T\mathcal{M}$. Also here, since $D_p$ is assumed to be possibly degenerate, $D_p^\sharp$ is in general non-invertible.
\end{itemize}

With reference to the time evolution equation \eqref{DE}, the GENERIC construction addresses both the non-dissipative (Hamiltonian) vector field $X^H$ and the dissipative vector field $Y^S$.

The Hamiltonian vector field $X^H_{\alpha(t)}$ is assumed to obtain from applying the Poisson operator $P^\sharp$ to the differential of the smooth Hamiltonian functional $H \colon \mathcal{M} \rightarrow \mathbb{R}$,
\begin{equation}\label{HamiltonianVectorField}
X^H_{\alpha(t)} = P_{\alpha(t)}^\sharp\left(\di H_{\alpha(t)}\right) \ ,
\end{equation}
while the dissipative vector field $Y^S_{\alpha(t)}$ is assumed to obtain from applying the friction operator $D^\sharp$ to the differential of the smooth entropy functional $S \colon \mathcal{M} \rightarrow \mathbb{R}$,
\begin{equation}\label{DissipativeVectorField}
Y^S_{\alpha(t)} = D_{\alpha(t)}^\sharp\left(\di S_{\alpha(t)}\right)\ ,
\end{equation}
subject to the following supplementary conditions.
 \begin{itemize}[leftmargin=*]
\item
The entropy functional $S$ must be chosen among the distinguished functionals (Casimir functionals) of the Poisson structure, i.e., the operator $P^\sharp$ must be such that
\begin{gather}
\{S, A\} = P(\di S, \di A) = \di A \left(P^\sharp(\di S)\right) = 0 \quad \forall A \in C^\infty(\mathcal{M}) \  ,\nonumber\\
\text{or, equivalently,} \quad P_{\alpha(t)}^\sharp \!\!\left(\di S_{\alpha(t)}\right) = 0 .\label{DegHam}
\end{gather}
\item  The Hamiltonian functional $H$ must be chosen among the distinguished functionals of the dissipative structure, i.e., the operator $D^\sharp$ must be such that
\begin{gather}
[H, A] = D(\di H, \di A) = \di A (D^\sharp(\di H)) = 0 \quad \forall A \in C^\infty(\mathcal{M}) , \nonumber \\
\text{or, equivalently,} \quad D_{\alpha(t)}^\sharp \!\!\left(\di H_{\alpha(t)}\right) = 0 .\label{DegDiss}
\end{gather}
\item Also the other conserved properties of the system are kept constants by the dynamics and, therefore, must be distinguished functionals of both brackets.
 \end{itemize}

As a result of these assumptions, if $\alpha(t)$  satisfies \Eq\eqref{DE} and $A$ is a smooth functional on $\mathcal{M}$, we calculate the directional derivative of $A$ along the velocity vector $\dot{\alpha}(t)$ in the following way:
\begin{align}
& \dfrac{\di}{\di t} \!\left( A_{\alpha(t)} \right)= dA_{\alpha(t)}\!\left(\dot{\alpha}(t)\right) = \nonumber \\
&= dA_{\alpha(t)} \!\left(X^H_{\alpha(t)} \right)+dA_{\alpha(t)} \!\left( Y^S_{\alpha(t)}\right) \nonumber \\
&= dA_{\alpha(t)} \!\left(P_{\alpha(t)}^\sharp \!\left(dH_{\alpha(t)}\right) \right)+dA_{\alpha(t)} \!\left( D^\sharp_{\alpha(t)} \!\left(dS_{\alpha(t)}\right)\right) \nonumber \\
&= P_{\alpha(t)} \!\left(dH_{\alpha(t)},dA_{\alpha(t)} \right)
+D_{\alpha(t)} \!\left(dS_{\alpha(t)},dA_{\alpha(t)}\right) \nonumber \\
&= \{H, A\}_{\alpha(t)} + [S, A]_{\alpha(t)}
\end{align}
or, in more synthetic symbolic notation,
\begin{equation}
\dot{A} = \{H, A\} + [S, A] \ ,
\end{equation}
where, however, we emphasize that $\dot{A}$ denotes neither the total nor the partial derivative of $A$ with respect to time ($A$ is not directly a function of time), but only and precisely what is written above.

From the degeneracy conditions, one easily sees that, for $A = H$,
\begin{equation}
\dot{H} = 0 \ ,
\end{equation}
which reflects the conservation of energy for an isolated system; and, for $A = S$,
\begin{equation}
\dot{S} = [S, S] \geq 0 \ ,
\end{equation}
which reflects the principle of entropy non-decrease.

We note in passing that the expression
\begin{equation}
D_p^\sharp\left(\di S_p\right)
\end{equation}
is similar in form to \Eq\eqref{gradient_expl}. The difference stems from the degeneracy of the tensor field $D_p$, which prevents us from identifying the expression $D_p^\sharp\left(\di S_p\right) $ as the gradient vector; this is because the degeneracy prevents a one-to-one correspondence between covectors and vectors. Therefore, we shall refer to $D_p^\sharp\left(\di S_p\right)$ as the entropy \ql gradient\qr, in quotation marks. Later, we discuss a supplementary condition (see \Eq\eqref{Leibniz}) that allows us to associate to it the meaning of a proper \emph{(horizontal) gradient}.

\subsubsection{Inner product formulation}

If the manifold is a vector space equipped with an inner product, as in the SEA case, we define variational derivatives according to \Eq\eqref{gradientI}, and introduce the notation
\begin{align}
\breve{L} &\eqdef P^\sharp R^{-1} \ , \\
\breve{M} &\eqdef D^\sharp R^{-1} \ .
\end{align}
We then have, using \Eq\eqref{variationalderivative},
\begin{align}
X^H_{\alpha(t)} &= P_{\alpha(t)}^\sharp R^{-1} R \left(\di H_{\alpha(t)}\right) = \breve{L}|_{\alpha(t)} \left( \left. \f[H]{p} \right|_{\alpha(t)}\right) \ \mbox{and} \\
Y^S_{\alpha(t)} &= D_{\alpha(t)}^\sharp R^{-1} R \left(\di S_{\alpha(t)}\right) = \breve{M}|_{\alpha(t)}\left( \left. \f[S]{p} \right|_{\alpha(t)} \right)\ ,
\end{align}
and we recover the usual form of the GENERIC,
\begin{equation}
\dot{\alpha}(t) = \breve{L}|_{\alpha(t)} \left(\left. \f[H]{p} \right|_{\alpha(t)}\right) + \breve{M}|_{\alpha(t)} \left(\left. \f[S]{p} \right|_{\alpha(t)}\right) \ .
\end{equation}

\section{Discussion}\label{Discussion}
The reformulation of the SEA model in the language of differential geometry makes it more easily comparable to the GENERIC model.

First of all, since in the SEA model the reversible or transport part of the dynamics is not rationalized as in GENERIC, but only described case by case, we see that the Poisson structure may be fully imported from GENERIC to SEA without changes. Hence, we shall focus on the dissipative part, analyzing similarities and differences between the two models and highlighting the aspects not completely clear and deserving further analyses.

The following subsections are not meant to be sequential. They can also be read in another order.

\subsection{Original purposes of the two models}
In their original article \cite{GO97}, the authors declared the two main purposes of GENERIC:
\begin{enumerate}
\item to reproduce known equations of motion of known physical theories by casting them in a single abstract form;
\item to suggest new equations for new thermodynamic theories dealing with complex systems.
\end{enumerate}
The goal of the SEA method \cite{gpB14} applied to meso- and macroscopic systems was similar:
\begin{enumerate}
\item to show that a broad selection of known theoretical frameworks for the description of non-equilibrium thermodynamics at various levels of description can all be unified when viewed as implementations of the SEA principle;
\item to provide  rigorous mathematical formalization of the so-called Maximum Entropy Production (MEP) Principle, as an attempt to clarifying its meaning, scope and domain of validity;
\item to propose a formalization of known theories which reduces to the linear theories in the proximity of equilibrium, entailing Onsager reciprocity. Hence, showing that such theories are indeed SEA with respect to any metric that at equilibrium reduces to a generalized Onsager conductivity matrix.
\end{enumerate}
However, the original SEA formulation developed for a very speculative and controversial quantum thermodynamics framework, motivated by the additional fundamental goal  to provide a technically consistent connection between a wealth of heuristic discussions about the Second Law and the Arrow of Time in the 70's and 80's. The attempt consisted in constructing a dynamical theory compatible with the Lyapunov stability requirement suggested by the Hatsopoulos-Keenan statement of the second law of thermodynamics  \cite{HK65,GB05}  whereby the maximum entropy states must be the only equilibrium states of the dynamics that are conditionally stable in the technical sense defined in Ref. \cite{gpB86b}, as already discussed above.

\subsection{Differences in the geometric structures}\label{geometry}
In the GENERIC model, the degeneracy condition \eqref{DegDiss} imposed on the geometrical structure of the manifold implies that every distinguished functional of the dissipative bracket cannot vary along the dissipative vector field. The entropy \qsl gradient\qsr $D^\sharp(\di S)$ is automatically parallel to the level sets of all the distinguished functionals (thus, for example, to the level sets of energy), i.e.,
\begin{equation}
\di A(Y^S) = \di A \!\left(D^\sharp(\di S)\right) = D(\di S, \di A) = [S, A] = 0 ;
\end{equation}
for every distinguished functional $A$ of the dissipative bracket. In other words, the information of constancy of the conserved functionals is contained already in the co-metric tensor $D$.

The SEA model adopts a sub-Riemannian structure \cite{rM06}. The conserved properties are constant on the submanifolds where purely dissipative time evolution takes place, and the condition \eqref{defTpMc}
defines a \emph{distribution} $\mathcal{D}$ \cite{foot3} through
\begin{equation}\label{distribution}
\mathcal{D}_p = \ker (\di c^1_p) \cap \ker(\di c^2_p) \cap \ldots \cap \ker(\di c^k_p) \, .
\end{equation}
Since the $\di c^i$'s need not necessarily be linearly independent everywhere, the distribution may be \emph{singular}. On this distribution one introduces a metric
\functiondef{g_p}{\mathcal{D}_p \times \mathcal{D}_p}{\mathbb{R}}{(u_p, v_p)}{g_p(u_p, v_p)}
with $g_p(u_p, v_p)>0$ for any nonzero $u_p$ and $v_p$.

Hereafter, we shall consider finite-dimensional manifolds, because the theorems we will mention are valid for this case. The distribution \eqref{distribution} is \emph{integrable}, i.e., we can find -- for each point $p$ -- an \emph{integral} submanifold $\mathcal{M}_{\cc}$ containing $p$ and such that $T_p\mathcal{M}_{\cc} = \mathcal{D}_p$ (see, e.g., \cite{jL13}): they are the intersections of the level sets of the conserved functionals. That is why, in paragraph \ref{SEAgeneralized}, we decided to give the \qsl more naive\qsr viewpoint, focusing on the integral submanifolds of the distribution rather than the distribution itself.

The geometric structure of the dissipative dynamics of metriplectic manifolds is also similar to a sub-Riemannian structure, but the distribution $D^\sharp(T^*\mathcal{M})$ is not necessarily integrable.

If, however, we impose the condition
\begin{equation}\label{condition}
\ker(D^\sharp) = \vspan(\{\di c^i\}) \ ,
\end{equation}
we return to the previous situation. This condition means that all the degeneracies in the GENERIC model are related to conservation laws.  In this case, as we will see in Section \ref{seaVSgeneric}, the SEA and GENERIC formalisms are perfectly equivalent.

The GENERIC model is more similar in spirit to Classical Mechanics, where the integrals of motion are generally unknown, and finding them is often a great challenge. In the SEA framework, conversely, knowing them is essential for the geometrical construction itself.

\begin{figure}[h!]
\centering
\includegraphics[width=1\linewidth]{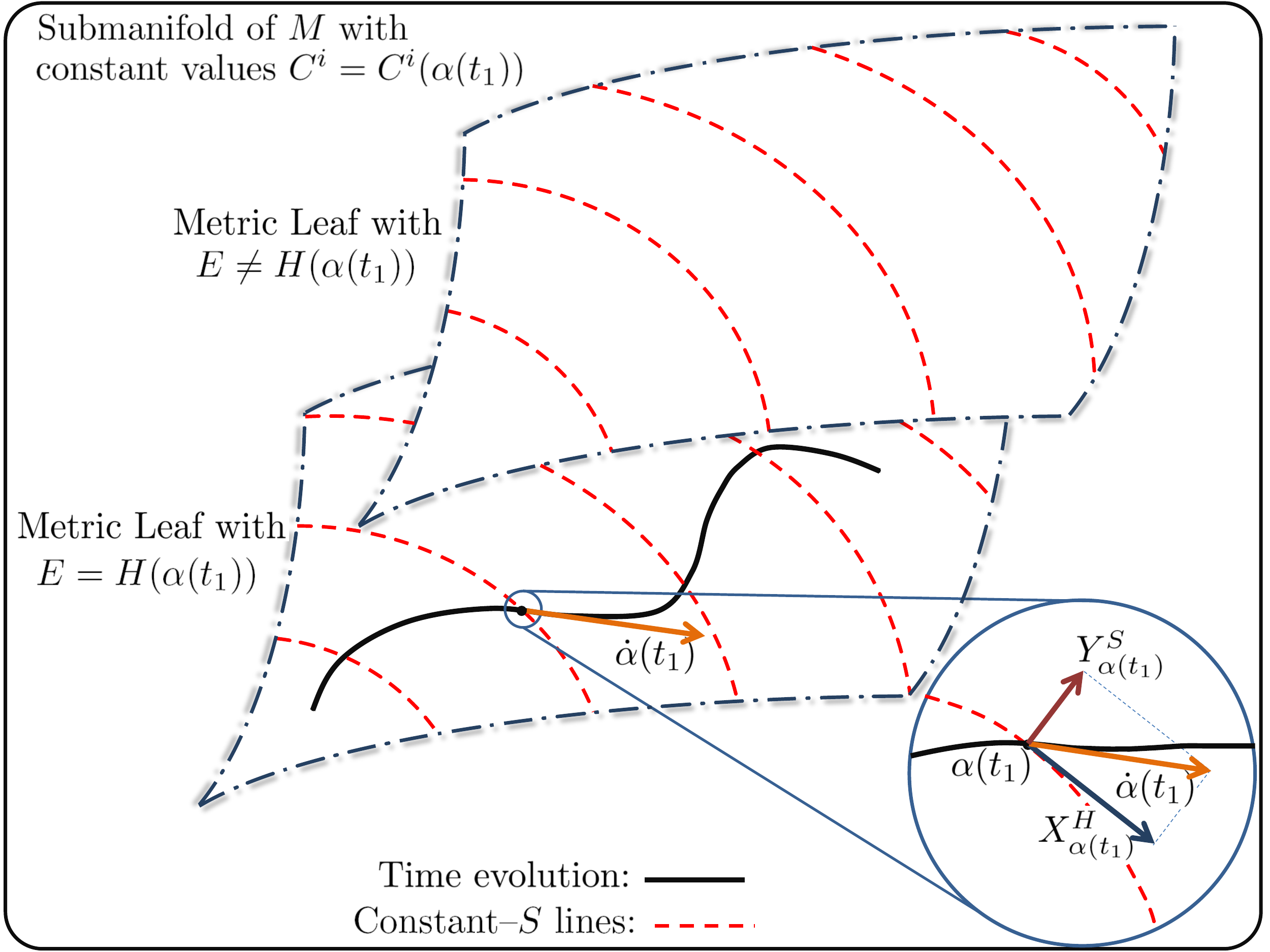}
\caption{Metric leaves in a manifold: GENERIC dynamics takes place on a single metric leaf.}
\label{Metric Leaf Picture}
\end{figure}

A pictorial visualization of the difference between the two constructions can be obtained by introducing in GENERIC the concepts of  \emph{symplectic leaves} and \emph{metric leaves}. Symplectic leaves are the submanifolds on which purely Hamiltonian evolution would take place (i.e., assuming  the dissipative vector  $Y^S$  set identically to zero) \cite{noteSymplectic}.  Metric leaves are instead the submanifolds where purely dissipative dynamics would take place (i.e., assuming the reversible vector $X^H$ set identically to zero). In the context of GENERIC dynamics, the degeneracy condition \eqref{DegHam} implies that symplectic leaves are at constant entropy (and the other distinguished  functionals  of the Poisson bracket) while the degeneracy condition \eqref{DegDiss} implies that metric leaves are at constant energy (and the other distinguished  functionals  of the dissipative bracket). Moreover, for an overall closed and isolated thermodynamic system, the time-independent Hamiltonian functional  is a constant of motion and, therefore, both the reversible vector $X^H$ and the dissipative vector  $Y^S$  lie in a metric leaf. Hence, the GENERIC dynamics cannot leave a particular metric leaf: each trajectory is effectively constrained on a single metric leaf, as shown pictorially in Figure \ref{Metric Leaf Picture}.

\begin{figure}[h!]
\centering
\includegraphics[width=1\linewidth]{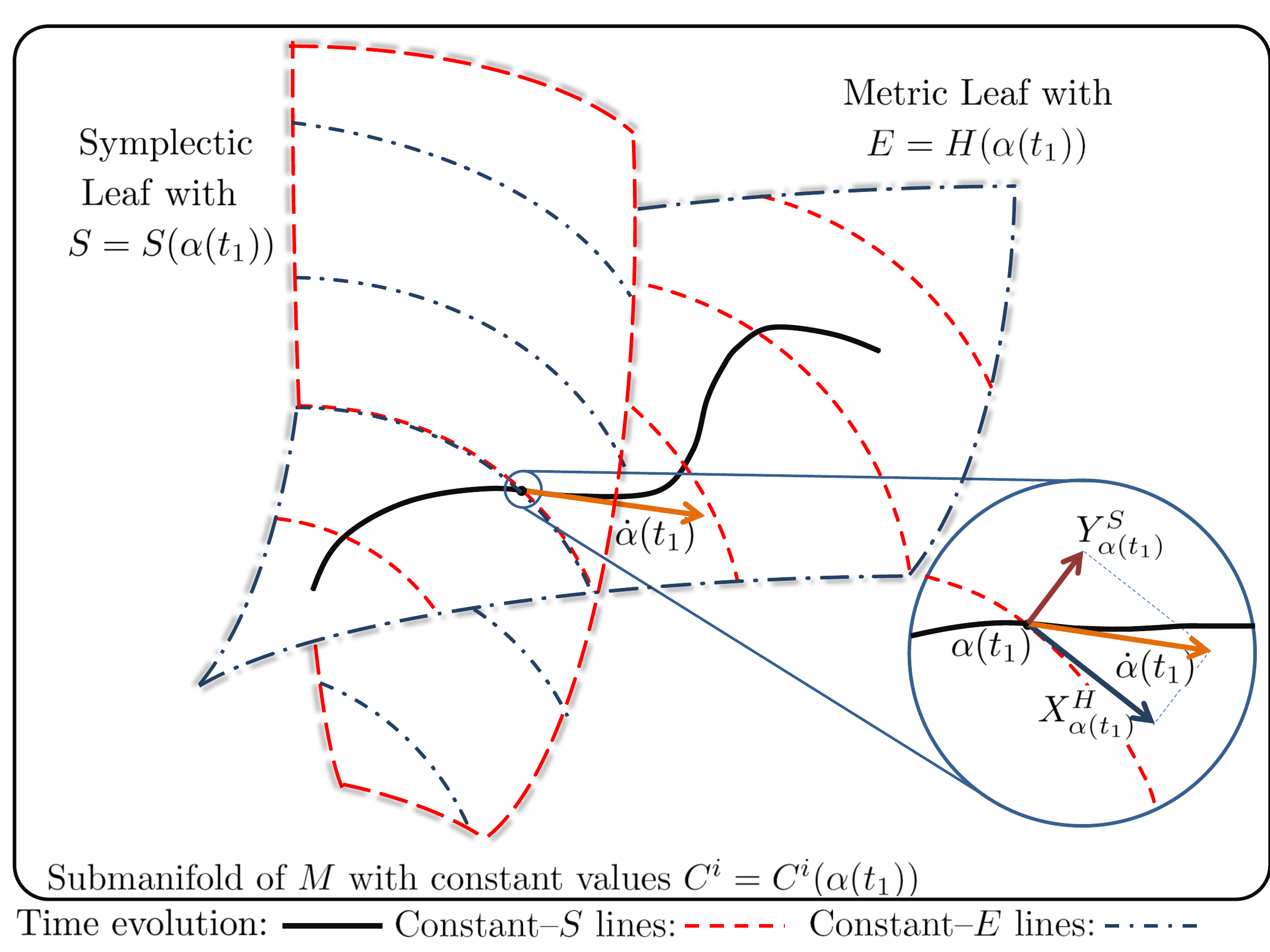}
\caption{The GENERIC reversible vector $X^H$ lies in the intersection of the metric leaf of the time evolution of an isolated system with the symplectic leaf (isentropic surface) being crossed at time $t_1$ by the time evolution.}
\label{Symplectic Metric Leaves}
\end{figure}

Figure \ref{Symplectic Metric Leaves} illustrates the relationship between metric leaves, where GENERIC dynamics (of an overall closed and isolated thermodynamic system) takes place, and symplectic leaves, where purely Hamiltonian dynamics takes place. Metric leaves are surfaces with constant energy, while the symplectic leaves are surfaces with constant entropy (because Hamiltonian dynamics is reversible). As a consequence, the intersection of symplectic leaves on a metric leaf produces isentropic contours and the GENERIC reversible vector $X^H$ (for an overall closed and isolated thermodynamic system) is always contained in such intersection.

In the SEA picture the situation is different for two reasons. The first is that the SEA model is not meant to be restricted  to the modeling of overall closed and isolated systems. Therefore,  the condition imposed within GENERIC that the vector field $X^H$ must be preserving the overall entropy is not necessarily imposed. Of course, for an isolated system the SEA construction can be made GENERIC by imposing the corresponding degeneracy onto the metric tensor. But the SEA model is meant to apply also for the description  of a continuum subjected to general boundary conditions.  By assuming a local state description instead of a global one, in terms of the local entropy density functional $s(\gamma)$ and the local (Lagrangian) entropy flux $\textbf{J}_{S}(\gamma)$, it allows one  to write the local entropy balance equation in the usual form
\begin{equation}\frac{\partial s}{\partial t} +\nabla\cdot\textbf{J}_S = \Pi_{S}  \label{Sbalance} \end{equation}
where $\nabla\cdot\textbf{J}_S$ represents the net rate of entropy outflow due to entropy-exchanging interactions between the local element of continuum and its neighbors, and of course it cannot be assumed equal to zero, due to the presence  in general of convective and diffusive fluxes of entropy, whereby also its volume integral is in general nonzero.
The local entropy production density $\Pi_{S}$ (often denoted by $\sigma$ in non-equilibrium thermodynamics) is controlled by the dissipative vector field $Y^S$ while the transport term $\nabla\cdot\textbf{J}_S$  is controlled by the reversible vector field $X^H$ (which of course controls also the local fluxes of other properties such as mass, momentum, angular momentum, energy). In order for GENERIC to be extended so as to include this class of descriptions, the degeneracy condition \eqref{DegHam} must be relaxed so as to properly account for mass, momentum, angular momentum, energy, and entropy fluxes across the boundaries of the system.  Some progress in this direction has been already made in Refs.\ \cite{mG93,hcO06}.

This version of the GENERIC (without imposing \Eq\eqref{condition})  could be made more similar to the SEA spirit of entropy production maximization by imposing an additional restriction on the GENERIC structure, as we point out below.

As is known from Poisson geometry, where Poisson manifolds foliate into symplectic leaves, a generalized distribution is integrable if and only if it is generated by a family of smooth vector fields, and is invariant with respect to their flows. This is the statement of the Stefan-Sussmann Theorem \cite{hjS73,pS74}, which is a generalization for singular distributions of the famous Frobenius Theorem. In Classical Mechanics, the condition that assures this integrability is the Jacobi identity, since it forces Hamiltonian flows to be canonical transformations (Poisson maps), that is, to preserve the Poisson structure.

In the GENERIC model, the time evolution of the state does not necessarily preserve the co-metric tensor, since dissipative flows themselves are not assumed to preserve the co-metric structure. For this reason, the distribution $D^\sharp(T^*\mathcal{M})$ is not integrable.

Instead, we note here that if we additionally endowed the dissipative structure with the \emph{Leibniz identity}
\begin{equation}\label{Leibniz}
\left[ \left[ A, B \right], C \right] = \left[ A, \left[ B, C \right] \right] + \left[ \left[ A, C \right], B \right] ,
\end{equation}
which is a generalization of Jacobi identity for non skew-symmetric brackets \cite{JacobyFootnote},   we would obtain that dissipative flows preserve the co-metric tensor, thus guaranteeing the integrability of the generalized distribution $D^\sharp(T^*\mathcal{M})$ to metric leaves. Then, on metric leaves, we would gain a (non-degenerate) metric, we could calculate distances with it, and we could define gradients by \Eq\eqref{gradient}. In this case, we could also interpret GENERIC dynamics as a SEA dynamics on metric leaves.

We may also express this result in general terms as follows. Suppose, for a moment, that there is only dissipative dynamics, so that time evolution is confined on a metric leaf in which the degenerate contravariant tensor $D$ is restricted into the non-degenerate one $D_L$. In this way, we can build the corresponding covariant metric tensor $g_L$, which acts on vectors as
\begin{equation}
g_L(u, v) = D_L \!\left(D_L^\flat(u), D_L^\flat(v)\right) \ .
\end{equation}
In finite dimension, it has matrix $[g_{L,ij}] = [D_L^{ij}]^{-1}$ (for a more rigorous treatment of this procedure for the case of symplectic leaves, see \cite{OR03}). Moreover, in the spirit of the variational formulation of the  SEA construction, let us consider all unit vectors $v$ at state $p$ ($g_L(v_p, v_p) = 1$) and search for the one that gives the maximal directional derivative of the entropy functional. By definition \eqref{gradient} of gradient of a smooth functional (given the non-degenerate bilinear form $g_L$) and the Cauchy-Schwarz inequality, we have
\begin{align}
\left| \di S_p(v) \right|^2 &= \left| g_L(\grad_L S|_p, v_p) \right|^2 \nonumber\\ &\leq
g_L\left(\grad_L S|_p, \grad_L S|_p\right)\ g_L(v_p, v_p) \nonumber\\ &= \left\| \grad_L S|_p \right\|_L^2 \ ,
\end{align}
that is, the absolute value of the directional derivative is always smaller than the norm of the gradient vector, and reaches its maximum value when
\begin{equation}v_p = \dfrac{\grad_L S|_p}{\left\| \grad_L S|_p \right\|_L} \ .\end{equation}
The restriction of the total entropy \qsl gradient\qsr $D_p^\sharp\left(\di S_p\right)$ to the metric leaf is indeed $\grad_L S|_p$, which is sometimes called \emph{horizontal gradient}.
Therefore, any nonequilibrium dynamics that can be written in GENERIC form and which satisfies the Leibniz identity \eqref{Leibniz} is automatically SEA on metric leaves.

However, while the Jacobi identity is a well-known feature of reversible dynamics with deep physical roots in Classical Mechanics, imposing an analogous condition on the dissipative structure is less founded on physical grounds.      For example, we leave it for further investigations to determine whether  \Eq \eqref{Leibniz} is satisfied and whether the  time evolution preserves the whole geometric structure in some of the mesoscopic thermodynamic formulations of dynamics  in which the state  description is in terms of a maximum entropy family. Such frameworks essentially adopt a kinematics compatible with some version of the maximum entropy formalism (see, e.g.,  Refs.\ \cite{rL86,GKZ04,Keck12}) whereby the nonequilibrium thermodynamic states are assumed to belong to the maximum entropy manifold defined by the local instantaneous values of a given set of mesoscopic properties chosen as the  \ql internal variables\qr of the system. Typically, this is the set of the constants of the motion augmented by a sufficient number of slowly varying additional properties characterizing some \ql constraints\qr (in the sense of Ref.\ \cite{Keck12}) or some \ql relevant information   that must be included in the analysis\qr (in the information theory sense, see, e.g., Ref.\ \cite{Caticha01,Caticha04,Caticha11}). For example, the recent Ref.\ \cite{mG14} assumes a maximum entropy kinematics of this kind and constructs on it a contact-structure-preserving dynamics  which effectively combines Hamiltonian and SEA dynamics. Again, Ref.\ \cite{gpB08} provides another example of maximal-entropy-generation model dynamics for a  maximum-entropy discrete-probability-distribution landscape with time-dependent constraints.

\subsection{Relaxation time  in SEA }\label{RelaxationTime}
Once the state space has been chosen, namely, the manifold where a thermodynamic process occurs, the SEA construction determines in a unique way the local direction of evolution at every point on the manifold and hence the trajectories of time evolution, but in order to do so it requires a choice for the notion of distance between states, i.e., more generally, the choice of a metric field on the tangent space to the submanifold with constant values of the conserved functionals.    This choice represents the \ql modeling knob\qr   that allows the description of different physical behaviors of the system. In other words, systems with identical kinematics and hence identical state spaces may exhibit different non-equilibrium dynamics: when in the same state they evolve differently. It is the metric field $g$ which characterizes the non-equilibrium behavior of the system. As argued in \cite{gpB14}, near equilibrium  the metric is directly related to the Onsager matrix of generalized resistances.

It is clear that the rate of evolution along the SEA trajectories is also regulated by the metric, since the velocity  of  a curve  parametrized by time $t$  is the scalar
\begin{equation}
\left\| \dot{\alpha}(t) \right\| = \sqrt{g(\dot{\alpha}(t), \dot{\alpha}(t))} ,
\end{equation}
which can be scaled by a constant in the metric tensor.

In the original quantum thermodynamic formulation of the SEA model, a metric (the Fisher-Rao metric) was chosen \textit{ab initio}: this was inspired by the fact that the state is, essentially, a probability measure and the interest was focused on identifying the simplest irreversible dynamics capable of incorporating the second law of thermodynamics.
The velocity along the SEA trajectory was scaled by the scalar $\tau$ which is allowed to be a functional of the state, i.e., to assume different values  along the trajectory in state space. In that simplest context $\tau$ represents also the single  relaxation time of the physical system being modeled. As shown in \cite[Eq. (87)]{gpB14}, $\tau$ can be interpreted as an \ql entropic time\qr because when  time $t$ is measured in units of $\tau$
the \ql speed\qr along the SEA trajectory is equal to the local rate
of entropy increase along the trajectory.  Moreover, $\tau$ is the Lagrange multiplier of the geometrical constraint in the  Lagrangian of the variational formulation of the SEA principle \cite[Eq. (72)]{gpB14}.  From the modeling point of view, $\tau$ is the only   knob to scale and control the \emph{strength of the attraction in the SEA direction} along the trajectories of the dynamics.

Here, however, we consider more general physical modeling contexts that include  the various frameworks explicitly considered in \cite{gpB14}, such as in complex chemical kinetics, or when multiple dissipative kinetic mechanisms give rise even far from equilibrium to Onsager-like couplings (like in thermodiffusion, thermoelectricity, etc.), or when isotropy is broken by the presence of phase interfaces or boundaries, or when preferential directions are imposed by externally applied fields. Then, the SEA model must account for the multiple  relaxation times in effect and this is obtained by assuming a non-isotropic (non Fisher-Rao) local metric tensor field, i.e., the operator $\hat{L} \eqdef \hat G^{-1} / \tau$ introduced in \cite{gpB14}, whose different eigenvalues represent the different local relaxation times up to a common scale factor.

In these more general contexts, the state dependence of the local metric tensor incorporates the information about the different kinetic mechanisms in act and their interplay, and fixes the ratios between all pairs of different relaxation times. The entropic time  $\tau$ could  be set to unity and thus absorbed in the metric tensor, but in the present work we prefer to show it explicitly for two reasons. The first is that  we wish to maintain a closer formal analogy between the structure of the SEA dissipative vector resulting from \Eq\eqref{SEAgram} and its equivalent in the original quantum thermodynamics framework. The second reason is to emphasize a somewhat philosophical  difference between the GENERIC and SEA approaches.

In fact, the choice considered more \ql natural\qr in GENERIC is to embed all the information about the dissipative part of the dynamics inside a single mathematical object, the friction operator. Instead, the \ql natural\qr choice in the SEA construction is to single out the three distinct geometrical aspects of the formalism by embedding them in three separate concepts: (1) the foliation of the state space induced by the constants of the motion; (2) the metric field that defines  the constrained entropy gradient needed to identify the SEA direction on the corresponding tangent space and, physically, incorporates the information about couplings and relaxation times of the different dissipative mechanisms in play;  and (3) the entropic time $\tau$ which regulates the speed with which the state evolves along the SEA path in state space.

\section{Boltzmann Equation}\label{BE}
In this section, we  illustrate how the two models are implemented in  Kinetic Theory, within the framework of validity of the Boltzmann Equation
\begin{equation} \begin{aligned}
& \p[f(\rr, \pp; t)]{t} =
\left[ \p[\phi(\rr)]{\rr} \cdot \p{\pp} - \dfrac{\pp}{m} \cdot \p{\rr} \right] f(\rr, \pp; t) \\
& \; + \iiint \di^3 p_2\iiint\di^3 q_1 \iiint \di^3 q_2 \,  w(\qq_1, \qq_2 | \pp, \pp_2) \\
& \quad \times \left[ f(\rr, \qq_1; t) f(\rr, \qq_2; t) - f(\rr, \pp; t) f(\rr, \pp_2; t) \right] \ ,
\end{aligned} \end{equation}
where $ f(\rr, \pp)$ is the one-particle distribution function, $\phi(\rr)$ is the potential of external forces, $w(\qq_1, \qq_2 | \pp_1, \pp_2)$ the transition probability given by
\begin{multline}
w(\qq_1, \qq_2 | \pp_1, \pp_2) =
\Dirac(\qq_1 + \qq_2 - \pp_1 - \pp_2) \\
\quad\times\Dirac(\qq_1^2 + \qq_2^2 - \pp_1^2 - \pp_2^2) \dfrac{8}{m}
\sigma(\qq_1, \qq_2 | \pp_1, \pp_2) \ ,
\end{multline}
and $\sigma$ the differential cross section calculated in the centre-of-mass frame. This is the formulation given by Grmela \cite{mG86} and \"Ottinger in \cite{hcO97}. The state space is the infinite-dimensional  vector space  $V$ of the distribution functions $f(\rr, \pp)$ that are well-defined, i.e., non-negative and with finite mean values of the meaningful moments.

At variance with Ref. \cite{hcO97},  we choose as suggested in Ref. \cite{gpB14} to reformulate the state description not in terms of  the distribution function $f$ but of  its square-root $\gamma(\rr, \pp)$ so that
\begin{equation}
f(\rr, \pp)= \gamma(\rr, \pp)^2   \ .
\end{equation}
This is done in order to accomplish three different scopes:
\begin{itemize}[leftmargin=*]
\item preserving the non-negativity of the distribution function;
\item making the gradients of the relevant physical properties belong to a Hilbert space $\mathcal{H}$;
\item avoiding the divergence of the entropy gradient outside of the support of $f(\rr, \pp)$, at the expense of an apparent singularity in the equation of motion.
\end{itemize}

The Boltzmann equation is recovered if the evolution equation for  $\gamma(\rr, \pp; t)$ is assumed of  the form
\begin{equation}\label{BoltzmannGamma} \begin{aligned}
& \!\!\!\! \p[\gamma(\rr, \pp; t)]{t} =
\left[ \p[\phi(\rr)]{\rr} \cdot \p{\pp} - \dfrac{\pp}{m} \cdot \p{\rr} \right] \gamma(\rr, \pp; t) \\
& \!\!\!\! +\frac{1}{2 \gamma(\rr, \pp, t)}
\iiint\di^3 q_1 \iiint    \di^3 q_2 \iiint \di^3 p_2 \, w(\qq_1, \qq_2 | \pp, \pp_2) \\
& \!\! \times \left[ \gamma(\rr, \qq_1; t)^2 \gamma(\rr, \qq_2; t)^2 - \gamma(\rr, \pp; t)^2 \gamma(\rr, \pp_2; t)^2 \right] .
\end{aligned} \end{equation}
\Eq\eqref{BoltzmannGamma} does present a divergence problem outside of the support of  $\gamma(\rr, \pp)$, but this is less problematic because the rates of change of all physical quantities depend on gradients which smooth out the divergence. In other words, they depend on  $\di\gamma^2/\di t$ which is free of this divergence issue.

 Each solution of the Boltzmann equation and of its thermodynamically consistent models is a one-parameter family of distribution functions  $\alpha \colon I \rightarrow \mathcal{H}$
where
\begin{equation}
\alpha(t) = \gamma(\rr, \pp; t)
\end{equation}
Thus, the Boltzmann equation and its GENERIC or SEA models take the abstract form of  the following differential equation
\begin{equation}\label{Boltzmann_abstract}
\dot{\alpha}(t) = X^H_{\alpha(t)} + Y^S_{\alpha(t)} ,
\end{equation}
where the explicit expressions of $X^H_{\alpha(t)}$ and $ Y^S_{\alpha(t)}$ differ in the GENERIC and the SEA approach as we have seen in the previous sections in abstract terms and we will see below in specific details for the present framework.

\subsection{GENERIC}\label{BEgeneric}

For the GENERIC construction, we consider the Hilbert space   $\mathcal{H}_\text{GENERIC} = L^2(\mathbb{R}^3\times\mathbb{R}^3)$ with inner product
\begin{equation}
\left\langle x,y \right\rangle = \iiint \di^3  r\iiint  \di^3 p\, x(\rr, \pp)\, y(\rr, \pp) \ .
\end{equation}
The overall mean values of the physical properties are represented by functionals $A[\gamma(\rr, \pp)]$ with associated local field $\tilde a(\rr, \pp,\gamma(\rr, \pp))$  such that  $\gamma(\rr, \pp)\,\tilde a(\rr, \pp,\gamma(\rr, \pp))$ belongs to $\mathcal{H}_\text{GENERIC}$. As a result,  the overall mean value functionals are
\begin{align}
A[\gamma(\rr, \pp)] &= \iiint \di^3 r\iiint  \di^3 p\, \gamma(\rr, \pp)^2 \tilde a(\rr, \pp,\gamma(\rr, \pp)) \nonumber\\ &= \left\langle \gamma , \gamma a \right\rangle =A \qquad \text{with} \quad |A| < \infty \ .
\end{align}
The normalization condition may be written as $I[\gamma(\rr, \pp)]=\left\langle \gamma, \gamma \right\rangle =1$.

 Since $\mathcal{H}_\text{GENERIC}$ is a vector space, every tangent space may be identified with the vector space itself, i.e., $T_p\mathcal{H}_\text{GENERIC} \cong \mathcal{H}_\text{GENERIC} \ \forall p$.
The functional derivative has the usual definition, analogous to Eq.\ \eqref{gradientI},
\begin{align}
& \left\langle\! {\left.\f[A]{\gamma}\right|_{\gamma_0}}, y \right\rangle = \di A_{\gamma_0}(y) \qquad \text{where} \\
&  \begin{cases} \gamma_0 \in \mathcal{H}_\text{GENERIC} \\ y \in T_{\gamma_0}\mathcal{H}_\text{GENERIC} \left( \cong \mathcal{H}_\text{GENERIC} \right) \end{cases} .
\end{align}

The fundamental properties which generate the dynamical equation in the GENERIC formulation are: the overall entropy functional
\begin{equation}
S[\gamma(\rr, \pp)] = - k_B \iiint \di^3 r\iiint  \di^3 p\, \gamma(\rr, \pp)^2 \ln\dfrac{\gamma(\rr, \pp)^2}{b}=S\  ,
\end{equation}
where $b$ is a suitable constant with the same dimensions as $\gamma^2$, and the overall mean value of the energy, which we write here below the other four collision invariant functionals representing the number of particles and the components of momentum
\begin{align}
& C^0[\gamma(\rr, \pp)] = \iiint \di^3 r\iiint  \di^3 p\, \gamma(\rr, \pp)^2 = N \ , \nonumber \\
& C^1[\gamma(\rr, \pp)] = \iiint \di^3 r\iiint  \di^3 p\, p_x\, \gamma(\rr, \pp)^2 = P_x \ , \nonumber \\
& C^2[\gamma(\rr, \pp)] = \iiint \di^3 r\iiint  \di^3 p\, p_y\, \gamma(\rr, \pp)^2 = P_y \ , \\
& C^3[\gamma(\rr, \pp)] = \iiint \di^3 r\iiint  \di^3 p\, p_z\, \gamma(\rr, \pp)^2 = P_z \ , \nonumber \\
& C^4[\gamma(\rr, \pp)] =
\iiint \di^3 r\iiint  \di^3 p \, \left[ \dfrac{\pp\cdot\pp}{2 m} + \phi(\rr) \right] \gamma(\rr, \pp)^2 = H \ . \nonumber
\end{align}
These can be rewritten in compact notation as
\begin{equation}
C^j[\gamma(\rr, \pp)] = \iiint\di^3 r \iiint \di^3 p\, \psi^j(\rr, \pp) \gamma(\rr, \pp)^2 = C^j\ ,
\end{equation}
where of course $\psi_0=1$, $\psi_1=p_x$, $\psi_2=p_y$, $\psi_3=p_z$, $\psi_4=\pp\cdot\pp/2 m +\phi(\rr)$.

The expressions for the functional derivatives are
\begin{align}
& \left.\f[S]{\gamma}\right|_{\gamma(\rr, \pp)} = - 2 k_B \gamma(\rr, \pp) \left[ \ln \dfrac{\gamma(\rr, \pp)^2}{b} + 1 \right] \ , \label{entropy differential} \\
& \left.\f[C^j]{\gamma}\right|_{\gamma(\rr, \pp)} = 2 \gamma(\rr, \pp)\ \psi^j(\rr, \pp) \ . \label{c differential}
\end{align}
If the state were chosen to be the distribution function $f(\rr, \pp)$, the expression \eqref{entropy differential} would present a divergence for values of $\rr$ and $\pp$ outside the support of the distribution function.

The functionals
\begin{equation}
C^iC^j[\gamma(\rr, \pp)] = \frac{1}{4}\left\langle\! \left.\f[C^i]{\gamma}\right|_{\gamma(\rr, \pp)},\left.\f[C^j]{\gamma}\right|_{\gamma(\rr, \pp)}\right\rangle =C^iC^j
\end{equation}
represent the overall mean values of the collision invariants for $i=0$ or $j=0$ and their overall moments otherwise.

The results in the rest of this subsection are borrowed from Ref. \cite{hcO97}, simply recast in terms of $\gamma(\rr, \pp)$ instead of $f(\rr, \pp)$  and written down in full detail.

In the abstract formulation of the GENERIC framework, the evolution equation takes the form
\begin{equation}
\dot{\alpha}(t) = X^{H,\text{GENERIC}}_{\alpha(t)} + Y^{S,\text{GENERIC}}_{\alpha(t)}
\end{equation}
where
\begin{align}
X^{H,\text{GENERIC}}_{\alpha(t)} &= \left.P_\gamma^\sharp\right|_{\alpha(t)} \left( \di H_{\alpha(t)} \right)
 = \left. \breve{L}_\gamma \right|_{\alpha(t)} \!\!\left(\! \left. \f[H]{\gamma} \right|_{\alpha(t)} \right)  \ , \\
Y^{S,\text{GENERIC}}_{\alpha(t)} &= \left.D_\gamma^\sharp\right|_{\alpha(t)} \left( \di S_{\alpha(t)} \right)
 = \left. \breve{M}_\gamma \right|_{\alpha(t)} \!\!\left(\! \left. \f[S]{\gamma} \right|_{\alpha(t)} \right)  \ .
\end{align}
More explicitly, the evolution equation for $\gamma(\rr, \pp; t)$ is
\begin{multline}
\p[\gamma(\rr, \pp; t)]{t} =   \left. \breve{L}_\gamma \right|_{\gamma(\rr, \pp; t)} \!\!\left(\! \left. \f[H]{\gamma} \right|_{\gamma(\rr, \pp; t)} \right) \\
+  \left. \breve{M}_\gamma \right|_{\gamma(\rr, \pp; t)} \!\!\left(\! \left. \f[S]{\gamma} \right|_{\gamma(\rr, \pp; t)} \right)  \ .
\end{multline}
We use the additional subscript $\gamma$ in $P_\gamma^\sharp|_{\alpha(t)}$, $ D_\gamma^\sharp|_{\alpha(t)}$,  $\breve{L}_\gamma |_{\alpha(t)}$ and $\breve{M}_\gamma |_{\alpha(t)}$  to distinguish these operators from the more standard ones that we give below in terms of $f=\gamma^2$, that we will denote by  $ P_f^\sharp|_{f(\rr, \pp; t)}$, $ D_f^\sharp|_{f(\rr, \pp; t)}$,  $\breve{L}_f |_{\alpha(t)}$ and $\breve{M}_f |_{\alpha(t)}$ .

\begin{widetext}
The Poisson operator at point $\gamma(\rr, \pp)$ is given by
\begin{equation}\label{Poisson}
 \left. \breve{L}_\gamma \right|_{\gamma(\rr, \pp)} \!\!\left(\! \left. \f[A]{\gamma} \right|_{\gamma(\rr, \pp)} \right)  = \dfrac{1}{2 \gamma(\rr, \pp)} \left[
\p{\pp}\,  \gamma(\rr, \pp)^2 \cdot \p{\rr}  -
\p{\rr}\,  \gamma(\rr, \pp)^2 \cdot \p{\pp}
\right] \!\left( \dfrac{1}{2 \gamma(\rr, \pp)} \left. \f[A]{\gamma} \right|_{\gamma(\rr, \pp)} \right) , \\
\end{equation}
and the associated Poisson bracket at point $\gamma(\rr, \pp)$
\begin{align}
& \{A, B\}_{\gamma(\rr, \pp)} = \left.P_\gamma\right|_{\gamma(\rr, \pp)}\!\!\left(
\di A_{\gamma(\rr, \pp)} , \di B_{\gamma(\rr, \pp)}\right) =  \di B_{\gamma(\rr, \pp)} \left(
\left.P_\gamma^\sharp\right|_{\gamma(\rr, \pp)} \!\!\left( \di A_{\gamma(\rr, \pp)} \right) \right) =
\left\langle \left.\f[B]{\gamma}\right|_{\gamma(\rr, \pp)} , \left. \breve{L}_\gamma \right|_{\gamma(\rr, \pp)} \!\!\left(\! \left. \f[A]{\gamma} \right|_{\gamma(\rr, \pp)} \right) \right\rangle  \nonumber \\
&= \frac{1}{4}\iiint \di^3 r\iiint  \di^3 p\, \dfrac{1}{ \gamma(\rr, \pp)}{\left.\f[B]{\gamma}\right|_{\gamma(\rr, \pp)}} \left[
\p{\pp}\,  \gamma(\rr, \pp)^2 \cdot \p{\rr}  -
\p{\rr}\,  \gamma(\rr, \pp)^2 \cdot \p{\pp}
\right] \!\left( \dfrac{1}{ \gamma(\rr, \pp)}\, \left.\f[A]{\gamma}\right|_{\gamma(\rr, \pp)} \right)
\nonumber \\
& =
\resizebox{\hsize}{!}{$ \displaystyle
\frac{1}{4}
 \iiint \di^3 r\iiint  \di^3 p\, \gamma(\rr, \pp)^2 \left[ \p{\rr}\! \left(\! \dfrac{1}{ \gamma(\rr, \pp)}{\left.\f[A]{\gamma}\right|_{\gamma(\rr, \pp)}}\right) \cdot \p{\pp} \! \left(\dfrac{1}{ \gamma(\rr, \pp)}{\left.\f[B]{\gamma}\right|_{\gamma(\rr, \pp)}}\right) \right. -
 \left.
\p{\pp}\! \left(\! \dfrac{1}{ \gamma(\rr, \pp)}{\left.\f[A]{\gamma}\right|_{\gamma(\rr, \pp)}}\right) \cdot \p{\rr}\! \left(\! \dfrac{1}{ \gamma(\rr, \pp)}{\left.\f[B]{\gamma}\right|_{\gamma(\rr, \pp)}}\right) \right]
 $}
 \nonumber \\
& =
\iiint \di^3 r\iiint  \di^3 p\, \gamma(\rr, \pp)^2\, \left[ \p{\rr} \left(\! \left.\f[A]{f}\right|_{f=\gamma(\rr, \pp)^2}\right) \cdot \p{\pp} \left(\! \left.\f[B]{f}\right|_{f=\gamma(\rr, \pp)^2}\right) \right. -
 \left.
\p{\pp} \left(\! \left.\f[A]{f}\right|_{f=\gamma(\rr, \pp)^2}\right) \cdot \p{\rr} \left(\! \left.\f[B]{f}\right|_{f=\gamma(\rr, \pp)^2}\right) \right]
  \nonumber \\
& =\{A, B\}_{f(\rr, \pp)=\gamma(\rr, \pp)^2} \ .
\end{align}
\end{widetext}

The friction operator at point $\gamma(\rr, \pp)$ can be written as follows
\begin{multline}\label{frictionGENERIC}
 \left. \breve{M}_\gamma \right|_{\gamma(\rr, \pp; t)} \!\!\left(\! \left. \f[A]{\gamma} \right|_{\gamma(\rr, \pp; t)} \right)  \\
= \iiint \di^3 p_1\, \hat{M}_\gamma[\gamma(\rr, \pp)](\rr, \pp, \pp_1) \left.\f[A]{\gamma}\right|_{\gamma(\rr, \pp_1)} \ ,
\end{multline}
where
\begin{align}\label{MGENERIC}
& \! \hat{M}_\gamma[\gamma(\rr, \pp)](\rr, \pp,  \pp_1)  \nonumber\\
& \!\! = \frac{1}{4k_B\gamma(\rr, \pp)\gamma(\rr, \pp_1)}\iiint \di^3 q_1 \iiint\di^3 q_2 \iiint\di^3 p_2 \, w(\qq_1, \qq_2 | \pp, \pp_2) \nonumber\\
& \, \times
 \resizebox{\hsize}{!}{$ \displaystyle
\left[ \Dirac(\pp - \pp_1) + \Dirac(\pp_2 - \pp_1) - \Dirac(\qq_1 - \pp_1) - \Dirac(\qq_2 - \pp_1) \right]
$}
\nonumber\\
& \, \times
 \dfrac{\gamma(\rr, \qq_1)^2 \gamma(\rr, \qq_2)^2 - \gamma(\rr, \pp)^2 \gamma(\rr, \pp_2)^2}
{\ln[\gamma(\rr, \qq_1)^2 \gamma(\rr, \qq_2)^2] - \ln[\gamma(\rr, \pp)^2 \gamma(\rr, \pp_2)^2]} \ ,
\end{align}
and the associated dissipative bracket at point $\gamma(\rr, \pp)$ reads
\begin{align}
& [A, B]_{\gamma(\rr, \pp)} =  \left.D_\gamma\right|_{\gamma(\rr, \pp)}
\!\!\left( \di A_{\gamma(\rr, \pp)} , \di B_{\gamma(\rr, \pp)} \right)  \nonumber \\
& =  \di B_{\gamma(\rr, \pp)}\left(\left. D_\gamma^\sharp\right|_{\gamma(\rr, \pp)} \left( \di A_{\gamma(\rr, \pp)}  \right)\right)  \nonumber \\
& = \left\langle \left.\f[B]{\gamma}\right|_{\gamma(\rr, \pp)} , \left. \breve{M}_\gamma \right|_{\gamma(\rr, \pp)} \!\!\left(\! \left. \f[A]{\gamma} \right|_{\gamma(\rr, \pp)} \right) \right\rangle  \nonumber \\
& =
\iiint \di^3 r\iiint  \di^3 p\, {\left.\f[B]{\gamma}\right|_{\gamma(\rr, \pp)}}  \left. \breve{M}_\gamma \right|_{\gamma(\rr, \pp)} \!\!\left(\! \left. \f[A]{\gamma} \right|_{\gamma(\rr, \pp)} \right)  \nonumber \\
& =
\resizebox{.95\columnwidth}{!}{$ \displaystyle
\iiint \di^3 r\iiint  \di^3 p\iiint \di^3 p_1\, {\left.\f[B]{\gamma}\right|_{\gamma(\rr, \pp)}}  \, \hat{M}_\gamma[\gamma(\rr, \pp)](\rr, \pp, \pp_1)\, {\left.\f[A]{\gamma}\right|_{\gamma(\rr, \pp_1)}}
$}
\nonumber \\
& =
\resizebox{.95\columnwidth}{!}{$ \displaystyle
\iiint \di^3 r\iiint  \di^3 p\iiint \di^3 p_1\, {\left.\f[B]{f}\right\rvert_{f(\rr, \pp)}}  \, \hat{M}_f[f(\rr, \pp)](\rr, \pp, \pp_1)\,{\left.\f[A]{f}\right|_{f(\rr, \pp_1)}}
$}
\nonumber \\
&=  \left\langle \left.\f[B]{f}\right|_{f(\rr, \pp)} , \left. \breve{M}_f \right|_{f(\rr, \pp)} \!\!\left(\! \left. \f[A]{f} \right|_{f(\rr, \pp)} \right) \right\rangle   \nonumber \\
&= [A, B]_{f(\rr, \pp)} \ ,
\end{align}
where we identify $\hat{M}_f$ as the dissipative \qsl matrix\qsr given in \Eq (12) of Ref. \cite{hcO97},
\begin{align}
& \! \hat{M}_f[f(\rr, \pp)](\rr, \pp,  \pp_1) \nonumber\\
& \!\! = \frac{1}{k_B}\iiint \di^3 q_1 \iiint\di^3 q_2 \iiint\di^3 p_2\,  w(\qq_1, \qq_2 | \pp, \pp_2) \nonumber\\
& \, \times \resizebox{.95\columnwidth}{!}{$ \displaystyle
\left[ \Dirac(\pp - \pp_1) + \Dirac(\pp_2 - \pp_1) - \Dirac(\qq_1 - \pp_1) - \Dirac(\qq_2 - \pp_1) \right] $}
\nonumber\\
& \, \times  \dfrac{f(\rr, \qq_1) f(\rr, \qq_2) - f(\rr, \pp) f(\rr, \pp_2)}
{\ln[f(\rr, \qq_1) f(\rr, \qq_2)] - \ln[f(\rr, \pp) f(\rr, \pp_2)]} \ ,
\end{align}
and the corresponding friction operator is
\begin{multline}
 \left. \breve{M}_f \right|_{f(\rr, \pp; t)} \!\!\left(\! \left. \f[A]{f} \right|_{f(\rr, \pp; t)} \right)  \nonumber\\ \quad= \iiint \di^3 p_2\, \hat{M}_f[f(\rr, \pp)](\rr, \pp, \pp_2) \left.\f[A]{f}\right|_{f(\rr, \pp_2)} \ ,
\end{multline}

It is easy but important to verify that the degeneracy requirements
\begin{equation}\label{Mgammadegeneracy}
 \left. \breve{M}_\gamma \right|_{\gamma(\rr, \pp; t)} \!\!\left(\! \left. \f[C^j]{\gamma} \right|_{\gamma(\rr, \pp; t)} \right)  =0 \qquad \forall\, j
\end{equation}
are consequence of the symmetry property (invariance upon exchange of $\qq_1,\qq_2$ with $\pp_1,\pp_2$) of both the transition probabilities $w( \qq_1,\qq_2|\pp_1,\pp_2)$  and the positive semi-definite resistance \qsl matrix\qsr
\begin{multline}
\Gamma(  \qq_1,\qq_2|\pp_1,\pp_2) \\ =\dfrac{\ln[f(\rr, \qq_1) f(\rr, \qq_2)] - \ln[f(\rr, \pp_1) f(\rr, \pp_2)]}{f(\rr, \qq_1) f(\rr, \qq_2) - f(\rr, \pp) f(\rr, \pp_1)}
\end{multline}
whose form was suggested by the related work in \cite{SS87}  on chemical kinetics and in the present kinetic theory framework  can be interpreted as a resistance matrix due to the collisions from $q_1,q_2$ to $p_1,p_2$ and viceversa. Indeed, the entropy production rate can be written as
\begin{multline*}
\!\!\!\!\!\!\! \Sigma = k_B\iiint \di^3 q_1 \iiint\di^3 q_2 \iiint\di^3 p_1 \iiint\di^3 p_2\, w(\qq_1, \qq_2 | \pp_1, \pp_2) \\
\ \times \Gamma( \qq_1,\qq_2|\pp_1,\pp_2)    \left[f(\rr, \qq_1) f(\rr, \qq_2) - f(\rr, \pp_1) f(\rr, \pp_2)\right]^2 \ .\\
\end{multline*}

In the case of GENERIC, the effort has been to put the Boltzmann Equation in GENERIC form, so that the Poisson operator and the friction operator $\hat{M}$ have arisen from this procedure. The friction operators  $\hat{M}_\gamma$ and $\hat{M}_f$  given above lead exactly to collision integral of the Boltzmann Equation. In spite of the complexity of such operators, it is hoped that knowing their explicit forms may help identify kinetic models of the Boltzmann collision integral in the same spirit of the BGK model but capable of capturing more features of the collision dynamics and of given better approximations in the far non-equilibrium domain. Early attempts along these lines are discussed in \cite{GPBNH13}.

\subsection{SEA}\label{BEsea}
For the SEA construction, the Hilbert space is  $\mathcal{H}_\text{SEA} = L^2(\mathbb{R}^3)$ with  (local) inner product  is
\begin{equation}
\left( x|y \right)(\rr) = \iiint  \di^3 p\, x(\rr, \pp)\, y(\rr, \pp) ,
\end{equation}
the local densities of the physical properties are $\rr$-dependent  functionals $ a(\rr)[\gamma(\rr, \pp)]$ with associated underlying field $\tilde a(\rr, \pp,\gamma(\rr, \pp))$ such that, for each fixed $\rr$,  $\gamma(\rr, \pp)\,\tilde a(\rr, \pp,\gamma(\rr, \pp))$ belongs to $\mathcal{H}_\text{SEA}$. As a result,  the local density functionals are
\begin{align}
\!\!\! a(\rr)[\gamma(\rr, \pp)] &= \iiint  \di^3 p\, \gamma(\rr, \pp)^2 \,\tilde a(\rr, \pp,\gamma(\rr, \pp)) \nonumber\\ &= (\gamma|\gamma a) = a(\rr) \quad\ \text{with} \quad |\hat{a}(\rr)| < \infty \ .
\end{align}

The functional derivative has again the usual definition analogous to Eq.\ \eqref{gradient}, but on $\mathcal{H}_\text{SEA}$,
\begin{equation}
\left({\left. {\left. \f[a]{\gamma}\right|_{\gamma_0}}\right| y} \right) = \di  a_{\gamma_0}(y) \quad \text{with}
 \begin{cases} \gamma_0 \in \mathcal{H}_\text{SEA} \\ y \in T_{\gamma_0}\mathcal{H}_\text{SEA} \left( \cong \mathcal{H}_\text{SEA} \right) \end{cases}
\end{equation}

Clearly,
\begin{equation}
\left\langle A,B \right\rangle = \iiint  \di^3 r\, \left( a| b \right)(\rr) \ .
\end{equation}

For the SEA formulation the local properties which generate the dynamical equation are
the  local density and flux fields defined as follows
\begin{align}
& s(\rr)[\gamma(\rr, \pp)] = - k_B\! \iiint  \di^3 p\, \gamma(\rr, \pp)^2 \ln\dfrac{\gamma(\rr, \pp)^2}{b} = s(\rr)\ , \nonumber \\
& c^j(\rr)[\gamma(\rr, \pp)] = \iiint \di^3 p\, \psi^j(\rr, \pp)\, \gamma(\rr, \pp)^2 =c^j(\rr) \ ,\\
& \JJ_{C^j}(\rr)[\gamma(\rr, \pp)] = \iiint \di^3 p\, \psi^j(\rr, \pp)\,\pp\, \gamma(\rr, \pp)^2 =\JJ_{C^j}(\rr) \ . \nonumber
\end{align}

The expressions for the functional derivatives are
\begin{align}
& \left.\f[s]{\gamma}\right|_{\gamma(\rr, \pp)} = - 2 k_B \gamma(\rr, \pp) \left[ \ln \dfrac{\gamma(\rr, \pp)^2}{b} + 1 \right]\label{entropy differential_local} \\
& \left.\f[c^j]{\gamma}\right|_{\gamma(\rr, \pp)} = 2 \gamma(\rr, \pp)\ \psi^j(\rr, \pp) \ .\label{c_differential_local}
\end{align}
We note that the rhs of \Eqs(\ref{entropy differential}) and (\ref{entropy differential_local}) are identical, and the same for \Eqs(\ref{c differential}) and (\ref{c_differential_local}).

Here, the functionals
\begin{equation}
c^ic^j[\gamma(\rr, \pp)] = \frac{1}{4}\left(\! \left.{\left. \f[c^i]{\gamma}\right|_{\gamma(\rr, \pp)}}\right|{\left.\f[c^j]{\gamma}\right|_{\gamma(\rr, \pp)}} \right)=c^ic^j(\rr)
\end{equation}
represent the local mean values of the collision invariants for $i=0$ or $j=0$ and the local moments otherwise.

In the abstract formulation of the SEA model, the evolution equation takes the form
\begin{equation}
\dot{\alpha}(t) = X^{H,\text{SEA}}_{\alpha(t)} + Y^{S,\text{SEA}}_{\alpha(t)}
\end{equation}
where we recall that $\alpha(t) = \gamma(\rr, \pp; t)$, the transport vector field
\begin{equation}
X^{H,\text{SEA}}_{\alpha(t)} = -\frac{\pp}{m}\cdot \p[\gamma(\rr, \pp; t)]{\rr} + \frac{\partial\phi(\rr)}{\partial \rr} \cdot \p[\gamma(\rr, \pp; t)]{\pp}
\end{equation}
is prescribed and not \qsl derived\qsr as in GENERIC, whereas the dissipative vector field is derived from the gradients of the entropy density and the conserved densities,
\begin{align}
Y^{S,\text{SEA}}_{\alpha(t)} &= \dfrac{1}{\tau}  g_{\alpha(t)}^\sharp \!\!\left(\di s^{\cc (\alpha(t))}_{\alpha(t)} \right) \nonumber\\   &= \dfrac{1}{\tau}   \hat G^{-1} \!\!\left(\! {\left.\f[s]{\gamma}\right|^{\cc(\alpha(t))}_{\alpha(t)}} \right) \ .
\end{align}
The values of the (Lagrange multipliers) $\beta^j$'s are found by solving the following system of five algebraic equations (\Eq\eqref{systemforbeta})
\begin{multline}\label{betasystem}
\sum\limits_{j = 0}^4 \left\langle\!{\left.\f[c^j]{\gamma}\right|_{\alpha(t)}},
{\left.\f[c^i]{\gamma}\right|_{\alpha(t)}}\right\rangle\, \beta^j_{\alpha(t)} \\
= \left\langle\!{\left.\f[s]{\gamma}\right|_{\alpha(t)}},
{\left.\f[c^i]{\gamma}\right|_{\alpha(t)}}\right\rangle \quad  i \in [0, 4] \ .
\end{multline}

The metric tensor $g$ or the equivalent operator $\hat G$  that makes the SEA formulation coincide with  the full Boltzmann equation can be in principle obtained by starting from the expression of the GENERIC friction operator $\breve{M}^\text{GENERIC}$ defined by \Eqs \eqref{frictionGENERIC} and \eqref{MGENERIC}, which corresponds to the full Boltzmann collision integral. In fact, in  the next section we prove that given a GENERIC friction operator $\breve{M}$, the metric tensor $g$ identified by \Eq \eqref{GENERIC2SEA} yields the equivalent SEA formulation. In particular, such $g$ is proportional through a scaling dimensionality  constant $\tau$ to the inverse of the restriction of $\breve M$  to  $\ker(\breve M)^{\perp}$.  The challenge of deriving the explicit expression of such metric tensor $g$ is left for future work.

The subsequent effort in the SEA philosophy, is to find an appropriate metric tensor capable to model correctly and efficiently the collision integral of
the Boltzmann Equation in the same spirit of the traditional  Kinetic Models, such as BGK, ES-BGK, etc. that constitute good approximations near-equilibrium,  so as to extend their validity to the  far non-equilibrium domain.  The problem of identifying criteria for this kind of models  is still open. Recent numerical results \cite{GPBNH13} show that the choice of a uniform (Fisher-Rao) metric yields poor models in this framework; more precisely, although near equilibrium it is fully equivalent to the BGK model, in the far non-equilibrium regime it selects trajectories in state space that diverge from the direction of evolution actually chosen by the full  Boltzmann collision integral. It is hoped that the present analysis and perhaps Information Geometry could provide  hints to find a suitable metric for this purpose.

\section{Equivalence of SEA and GENERIC (in most frameworks)}\label{seaVSgeneric}
In this section we show that every SEA model admits a GENERIC form, of course, after making the choice of  a kinematics, which is the common starting point. In other words, we prove that we can construct the GENERIC form of any given SEA model. We also prove the converse to be true.

This result holds in the Kinetic Theory framework of validity of the Boltzmann Equation that we considered in the previous section for illustrative purposes. But they are also of much broader validity in that they hold at least  for all the frameworks for which the SEA constructions have been made explicit in Ref. \cite{gpB14}. To show such broader validity, below we state the result with explicit reference to the Kinetic Theory framework but by using a more compact notation which points directly to the notation introduced in \cite{gpB14} in order to unify several different non-equilibrium frameworks and levels of description. In particular, we introduce the following notation, giving a uniform treatment to the symbols used in the section regarding the SEA and GENERIC interpretations of the Boltzmann Equation.

Like in the previous sections, we use the same symbol $\gamma(\rr, \pp) $ to denote the states in GENERIC and SEA, even though in SEA the position $\rr$ is a fixed parameter also for the local functionals. What is important, though, is that the proper functional derivatives in the two frameworks end up being identical functions of   $\rr$ and $\pp$. Therefore, we denote them by the same symbol.
We write the functional derivative of entropy as
\begin{equation}
|\Phi)=\left.\f[s]{\gamma}\right|_{\gamma(\rr, \pp)}=\left.\f[S]{\gamma}\right|_{\gamma(\rr, \pp)} ,
\end{equation}
collect a complete set of  conserved quantities in the vectors
\begin{equation}
\cc=\{c^j\},\qquad \CC=\{C^j\} \ ,
\end{equation}
write their functional derivatives as
\begin{equation}
|\PPsi)=\left.\f[\cc]{\gamma}\right|_{\gamma(\rr, \pp)}=\left.\f[\CC]{\gamma}\right|_{\gamma(\rr, \pp)} \ ,
\end{equation}
and for simplicity, without loss of generality, assume they are linearly independent (otherwise we drop from sets $\cc$ and $\CC$ the conserved quantities that do not have independent functional derivatives, as discussed in Section \ref{SEAoriginal} after \Eq\eqref{systemforbeta}).

We use the GENERIC friction operator $\breve{M}$ (dropping the apex \qsl GENERIC\qsr), which acts on a vector $b$ on $T_\gamma \mathcal{H}$  according to \Eq\eqref{frictionGENERIC}:
\begin{equation}
\breve{M} \ :\qquad \breve{M} |b) = D_\gamma^\sharp(b^*) \ ,
\end{equation}
where $b^*$  is the corresponding covector (the two may be identified thanks to the presence of the inner product). For the SEA operators, we have, for $b^{\cc}$ on $T_\gamma \mathcal{H}$,
\begin{align}
& \hat{G} \ :\qquad  \hat{G}|b^{\cc}) = g_\gamma^\flat(b^{\cc}) \ , \\
& \hat{L} \ :\qquad  \hat{L}|b^{\cc}) = \frac{1}{\tau}\hat{G}^{-1}|b^{\cc}) = \frac{1}{\tau}g_\gamma^\sharp(b^*) \ ,\\
& ( a^{\cc}|\hat{G}|b^{\cc})=( a^{\cc},g_\gamma^\flat(b^{\cc}))=g_\gamma(a^{\cc},b^{\cc}) \ .
\end{align}

Finally, the dissipative part of the local dynamics, i.e., the part responsible for local entropy generation, like the Boltzmann collision integral in the Boltzmann Equation, and the Lagrange multipliers are:
\begin{align}
& |\Pi_\gamma) = Y^S_{\alpha(t)} \ , \\
& \bbeta=\{\beta^j_{\alpha(t)}\} \ , \\
& |\bbeta\cdot \PPsi) = \sum_j \beta^j_{\alpha(t)} \left.\f[c^j]{\gamma}\right|_{\gamma(\rr, \pp)} \ .
\end{align}

Within the GENERIC framework, $|\Pi_\gamma)$ takes the form
\begin{equation}\label{GENERICcompact}
|\Pi_\gamma) = \breve{M} |\Phi)
\end{equation}
where $\breve{M}$ is  subject to the conditions
\begin{equation}\label{GENERICconditions}
\breve{M} |\PPsi)=0 \quad , \quad \breve{M} \ge 0 \quad \text{and} \quad \breve{M} \text{ symmetric},
\end{equation}
whereas within the SEA framework it takes the form
\begin{equation}\label{SEAcompact}
|\Pi_\gamma) = \hat{L}|\Phi^{\cc})=\hat{L}|\Phi-\bbeta\cdot \PPsi)
\end{equation}
where $\hat{L}$ is  subject to the conditions
\begin{equation}\label{SEAconditions}
\hat{L} > 0 \quad \text{symmetric and defined on}\ \vspan(\PPsi)^\perp  \ ,
\end{equation}
and $\bbeta$ is given by (\Eq\eqref{betaDirect})
\begin{equation}\label{Lagrange}
\bbeta = (\PPsi|\PPsi)^{-1}\cdot (\PPsi|\Phi) \ ,
\end{equation}
where  $(\PPsi|\PPsi)^{-1} $ denotes the inverse of matrix $(\PPsi|\PPsi)$ with elements $[\langle\Psi^i,\Psi^j\rangle]$.

\subsection{GENERIC form of a SEA model}\label{SEA2GENERIC}

Now, to prove that every SEA model admits a GENERIC form, we note that  we can rewrite \Eq (\ref{SEAcompact}) as
\begin{equation}\label{SEAcompact2}
|\Pi_\gamma) = \hat{L}\hat P_{\vspan(\PPsi)^\perp}|\Phi) \ .
\end{equation}
Before we conclude that the operator
\begin{equation}\label{SEA2GENERIC}
\breve{M}_{\hat{L},\PPsi} = \hat{L}\hat P_{\vspan(\PPsi)^\perp}
\end{equation}
 provides the GENERIC form  (\ref{GENERICcompact}) of the SEA dynamical equation (\ref{SEAcompact}), we must show that
$\breve{M}_{\hat{L},\PPsi}$  satisfies the requirements stated in \Eq (\ref{GENERICconditions}). In fact,  the first condition is a consequence of $\hat P_{\vspan(\PPsi)^\perp} |\PPsi)=0$, from which it also follows that $\ker(\breve{M}_{\hat{L},\PPsi})=\vspan(\PPsi)$ and  when restricted to $\vspan(\PPsi)^\perp$ operator $\hat P_{\vspan(\PPsi)^\perp}$ is the identity and   $\breve{M}_{\hat{L},\PPsi}$ reduces to $\hat L$. The second and third conditions are direct consequences of the symmetry and positive definiteness of $\hat{L}$.  To prove even more explicitly that $\breve{M}_{\hat{L},\PPsi}$ is positive semi-definite, consider any vector $|b)$ in $T_\gamma \mathcal{H}$ and its decomposition $|b)=|b^{\cc})+|b^{\perp\cc})$ where $|b^{\cc})= \hat P_{\vspan(\PPsi)^\perp} |b)$ and $|b^{\perp\cc})=|b)-|b^{\cc})$. Then, we have
$(b|\breve{M}_{\hat{L},\PPsi}|b)
=(b^{\cc}+b^{\perp\cc}|\hat{L}\hat P_{\vspan(\PPsi)^\perp}|b^{\cc}+b^{\perp\cc})
=(b^{\cc}+b^{\perp\cc}|\hat{L}|b^{\cc})
=(b^{\cc}|\hat{L}|b^{\cc})
\ge 0$ with the equal sign holding only when $|b^{\cc})=0$, i.e., when $|b)$  lies in the kernel of $\vspan(\PPsi)$.

\Eq (\ref{SEA2GENERIC}) supports explicitly our assertion in Section \ref{RelaxationTime} that the GENERIC friction operator incorporates both the information about the  constants of the motion (it projects onto the local metric leaf orthogonal to their gradients) and the information about the local metric on such leaf so that when applied to the entropy gradient it essentially identifies the SEA direction compatible with the local conservation constraints.

This concludes the proof that any SEA formulation can always be put in GENERIC form. Therefore, all the frameworks discussed in \cite{gpB14}, once put into SEA form by choosing the suitable co-metric $\hat{L}$,  can also be put into GENERIC form (at least as regards the dissipative part)  by using the $\breve{M}$   given by \Eq (\ref{SEA2GENERIC}) with $\hat{L}=\hat{G}^{-1}/\tau$. In other words, for any operator $\hat{L}$, the operator $\breve{M}$ given in \Eq (\ref{SEA2GENERIC}) makes the rhs of \Eq (\ref{GENERICcompact}) become identical to the rhs of \Eq (\ref{SEAcompact}).

\subsection{SEA form of a GENERIC model}\label{GENERIC2SEA}

Next, we show that also the converse is true, i.e., that any GENERIC formulation can always be put into SEA form provided \Eq\eqref{condition} holds. To do that, given a GENERIC friction operator $\breve{M}$, we first identify its kernel $\ker(\breve{M})$ and then select as  constants of the motion for the SEA formulation a set of state functionals such that their functional derivatives $|\PPsi)$ form a basis for $\ker(\breve{M})$. This way the dissipative vector fields in both models will conserve the same state functionals. As a result of this choice, \begin{equation}\label{converseCondition1}
\ker(\breve{M}) = \vspan(\PPsi) ,
\end{equation}
where $\vspan(\PPsi)$ denotes the linear span of the set of vectors $|\PPsi)$. Clearly,
also the following condition holds:
\begin{equation}\label{converseCondition}
\hat{P}_{\ker(\breve{M})} = \hat{P}_{\vspan(\PPsi)} .
\end{equation}

In the framework of the Boltzmann equation, it is well known \cite{cC88, cC90} that the kernel of the collision integral coincides with the linear span of the five collision invariants $\psi_0=1$, $\psi_1=p_x$, $\psi_2=p_y$, $\psi_3=p_z$, $\psi_4=\pp\cdot\pp/2 m +\phi(\rr)$, i.e., there exist no other linearly independent collision invariants. Since the friction operator given by \Eq\eqref{frictionGENERIC} and \Eq\eqref{MGENERIC} has been proven to be exactly equivalent to the full Boltzmann collision integral, by applying it to the functional derivatives in \Eq\eqref{c differential}, it is easy to verify that the well known result implies that \Eq\eqref{converseCondition} holds for $\breve{M}_\gamma$.

 In other frameworks, condition \eqref{converseCondition} appears as a reasonable additional condition in every GENERIC construction, for otherwise the structure would admit more distinguished functionals of the dissipative structure than the actual conserved properties, i.e., in other words there would be some sort of \emph{hidden} additional constants of the motion.

 To proceed with the proof,  let us consider the operator $\hat{L}_{\breve{M}}$ defined by the restriction of $\breve{M}$ on $\ker(\breve{M})^\perp$, i.e.,
 \begin{equation}\label{GENERIC2SEA}
\hat{L}_{\breve{M}}|b^{\cc})= \breve{M}|b^{\cc})\quad \forall |b^{\cc}) \in \ker(\breve{M})^\perp\ .
\end{equation}
In view of the degeneracy requirements $\breve{M}|\PPsi)=0$,  operator $\hat{L}_{\breve{M}}$ is readily shown to convert the SEA \Eq\eqref{SEAcompact} into the GENERIC \Eq\eqref{GENERICcompact}. Indeed,
 \begin{equation}
\hat{L}_{\breve{M}}|\Phi^{\cc})= \breve{M}|\Phi^{\cc})=\breve{M}|\Phi-\bbeta\cdot\PPsi)=\breve{M}|\Phi)\ .
\end{equation}
 This concludes the proof that we can construct the SEA form of any given GENERIC model.

In order to identify the metric $\hat G=\hat L^{-1}/\tau$ which makes the Boltzmann Equation fit exactly into the SEA form, we would need to identify $\ker(\breve{M}_\gamma^\text{GENERIC})$ for the dissipative operator $\breve{M}_\gamma^\text{GENERIC}$ given by \Eq (\ref{MGENERIC}). We leave the task of finding
 the explicit expression of $\hat{P}_{\ker(\breve{M})}$ for future work.

\section{Conclusions}\label{Conclusions}
The main objective of the present paper is the comparison between the Steepest Entropy Ascent (SEA) dynamical model, initially proposed by Beretta in a quantum thermodynamics framework and recently adapted to meso- and macroscopic systems, and the GENERIC (General Equation for Non-Equilibrium Reversible-Irreversible Coupling) formalism, developed by Grmela and \"Ottinger. To this end, we reformulated the SEA formalism  using the notation of Differential Geometry similar to that already available for the GENERIC formalism.

Our detailed analysis shows  that the two non-equilibrium dynamical models show similar patterns in that both may be considered as belonging to the \emph{maximal-entropy-producing} or the \emph{entropy-gradient} type. In both models the irreversible component of the time evolution of the state of a thermodynamic system is determined by the differential of the entropy functional.  In the SEA model it is in the direction of the projection $\di S^{\cc}$ of $\di S$  onto the submanifold  where the conserved properties are constant. In the GENERIC model it is in the direction of the entropy \ql gradient\qr in the metric leaf corresponding to the constant values of the conserved properties (the reason we put gradient between quotation marks is explained at the end of section IIB).

Both structures have been motivated by the search for Non-Equilibrium Thermodynamics formulations that are fully compatible with the Second Law of Thermodynamics. However, specific differences must be pointed out:
\begin{itemize}[leftmargin=*]
\item The SEA construction focuses  only on the irreversible component of the dynamics, and describes it by assuming the existence of a sub-Riemannian metric tensor field.
\item The GENERIC construction tackles with equal emphasis both the reversible and the irreversible components of the dynamics, and assumes a Poisson structure to describe the non-dissipative component  and a degenerate co-Riemannian structure to describe the dissipative  component.
\item A SEA model requires the separate specification of: (1) a set of time-invariant state functionals $\cc(p)$ representing constants of the motion or constraints, whose variational derivatives $\PPsi$ determine at every state $p$ the tangent space $T_p\mathcal{M}_{\cc(p)}=\vspan({\PPsi_p})^\perp$ to the submanifold $ \mathcal{M}_{\cc(p)} $ that contains the dissipative component $Y^S_{\alpha(t;p)}$ of the equation of motion; (2) a metric field $\hat{G}_p$ which for every state $p$ in the state manifold $\mathcal{M}$ defines the geometric notion of distance on the constrained  submanifold $ \mathcal{M}_{\cc(p)} $. Physically, the metric tensor $\hat{G}_p$ extends the notion of  generalized Onsager resistivity  to the far-from-equilibrium domain.
\item The  dissipative part of a GENERIC model requires the specification of a degenerate operator $ \breve{M}$ on the space $T_p\mathcal{M}$ tangent to the state manifold $\mathcal{M}$. We have shown that when $ \breve{M}$ is constructed so that its kernel $\ker(\breve{M})$ coincides with the linear span of the functional derivatives $\PPsi$ of the  time-invariant state functionals  and  its restriction to $\ker(\breve{M})^\perp$ is non-negative definite and symmetric, then the model is essentially SEA.
\item The GENERIC friction operator $ \breve{M}$ incorporates both the information about the constants of the motion (it projects onto the local metric leaf orthogonal to their variational derivatives) and the information about the local metric on such leaf so that when applied to the entropy variational derivative it essentially identifies the SEA direction compatible with the conservation constraints. In other words, in the GENERIC formalism the conservation laws are embedded in the degeneracy of the two assumed geometrical structures, while the SEA formalism assumes that the conservation constraints are given explicitly so as to determine the submanifolds where the purely dissipative time evolutions would lie and unfold along the direction of SEA with respect to a metric. The metric represents the couplings and characteristic times of the different dissipative mechanisms in act.
\item In SEA dynamics, the choice of a non-degenerate metric allows one  to univocally define gradients, while in the GENERIC formalism, the choice of a degenerate metric makes it impossible to define a metric and, thus, a gradient, unless a further condition on the dissipative bracket is imposed.
\item For the description of a continuum, SEA dynamics emerges as a local theory that starts from the local balance equations and implements the assumption of maximal local entropy production density compatible with the local conservation constraints, while the GENERIC formalism emerges as a global theory that implements an entropy gradient dynamics compatible with the global conservation constraints.
\end{itemize}
Nevertheless, in this paper we show that the descriptions of the dissipative components of the dynamics in the two theories are very closely related, and in some important instances entirely equivalent.

This is the case, for example, of the Boltzmann Equation, that we work out explicitly in both frameworks not only for illustrative purposes but also to prove the new result that the already known GENERIC form of the collision integral can also be given a SEA form.
The two models  have emerged in Kinetic Theory with different motivations. On one hand, SEA dynamics -- which was originally developed \cite{sGS01add} as an attempt to understand the fundamental consequences of an attempt to construct a theory of  quantum thermodynamics by embedding the second law directly into quantum theory -- has been adapted to the framework of Kinetic Theory with the aim of
   finding a \emph{simplified} metric to model the collision integral \cite{GPBNH13}, in order to create efficient Kinetic Models capable of extending to the highly non-equilibrium regime traditional near-equilibrium models such as BGK and ES-BGK. On the other hand, GENERIC, according to one of the two purposes for which the model was developed by its authors, aims at proving that the Boltzmann Equation is a realization of their general abstract dynamics.

Some of the topics considered in the present paper are in need of further ideas or deserve a deeper analysis:
\begin{itemize}[leftmargin=*]
\item the reversible-part formalism that GENERIC borrows from Geometric Mechanics may be \qsl trasferred\qsr to SEA in order to have a more complete model that explicitly considers Hamiltonian dynamics;
\item the idea of imposing that the dissipative bracket in GENERIC satisfies the \emph{Leibniz identity} in order to have a non-degenerate metric on the metric leaves, could be tested in practical instances by a symbolic algorithm as done for the Jacobi identity in \cite{KHO01};
\item as far as open systems are concerned, a parallel could be undertaken between the approach used in \cite{gpB14} and the mathematical framework of Dirac structures, which the authors of GENERIC claim to play a role in this kind of modeling \cite{hcO05}.
\end{itemize}


\begin{thebibliography}{10}




\bibitem{lO31} L. Onsager, Reciprocal Relations in Irreversible Processes. I, Phys. Rev. \textbf{37}, 405--426 (1931).

\bibitem{gpB86b}
G.P. Beretta, A theorem on Liapunoff stability for dynamical systems and a
  conjecture on a property of entropy, J. Math. Phys. \textbf{27}, 305--308 (1986).

\bibitem{rM00}
R. Mrugala, On contact and metric structures on thermodynamic spaces, RIMS Kokyuroku (Kyoto University) \textbf{ 1142}, 167--181 (2000).


\bibitem{viA89}
V.I. Arnol'd, {\em Mathematical Methods of Classical Mechanics} (Springer-Verlag, 2 edition,  New York, 1989; 1st ed., Nauka, 1974).

\bibitem{MR03}
J.E. Marsden and T.S. Ratiu, {\em Introduction to Mechanics and Symmetry} (Springer-Verlag, New York, 2003; 1st ed., 1994).

\bibitem{cC09}
C. Carathéodory, Untersuchungen ueber die Grundlagen der Thermodynamik, {\em Mathematische Annalen} \textbf{67}, 355--386 (1909).

\bibitem{rH73}
R. Hermann, {\em Geometry, Physics and Systems} (Marcel Dekker, New York, 1973).

\bibitem{hQ07}
H. Quevedo, Geometrothermodynamics, J. Math. Phys. \textbf{ 48}, 013506 (2007).

\bibitem{pjM09}
P.J. Morrison, Thoughts on brackets and dissipation: Old and new, J. Phys. Conf. Series \textbf{169}, 012006 (2009).

\bibitem{dF05}
D. Fish, Metriplectic systems, Ph.D. thesis, Portland State University, 2005.


\bibitem{GO97}
M. Grmela and H.C. \"Ottinger, Dynamics and thermodynamics of complex fluids. I. Development of a   general formalism, Phys. Rev. E \textbf{56}, 6620--6632 (1997).


  \bibitem{gpB86}
G.P.  Beretta, Steepest entropy ascent in quantum thermodynamics, in {\em The Physics of Phase Space (Nonlinear Dynamics and Chaos, Geometric Quantization, and Wigner
  Function), Proceedings of the First International Conference on the Physics
  of Phase Space, University of Maryland, College Park, May 20-23, 1986},  edited by Y.S. Kim and W.W. Zachary (Springer-Verlag, New York, 1986),  pp. 441--443.


\bibitem{gpB87}
G.P. Beretta, Steepest-ascent constrained approach to maximum entropy, in  {\em Second Law Analysis of Heat   Transfer in Energy Systems}, edited by R.F. Boehm and N. Lior, ASME Book G00390, HTD \textbf{80}  31--38 (1987).


\bibitem{sGS01}
S. Gheorghiu-Svirschevski, Nonlinear quantum evolution with maximal entropy production, Phys. Rev. A  \textbf{63}, 22105 (2001).

\bibitem{sGS01add}
S. Gheorghiu-Svirschevski, Addendum to "Nonlinear quantum evolution with maximal entropy production", Phys. Rev. A  \textbf{63}, 054102 (2001).

\bibitem{gpB09}
G.P.  Beretta, Nonlinear quantum evolution equations to model irreversible adiabatic relaxation with maximal entropy production and other nonunitary processes, Reps. Math. Phys. \textbf{64}, 139--168 (2009).


 \bibitem{CBS15}
 S. Cano-Andrade, G.P. Beretta, and M.R. von Spakovsky, Steepest-entropy-ascent quantum thermodynamic modeling of decoherence in two different microscopic composite systems, Phys. Rev. A \textbf{91}, 013848 (2015).


  \bibitem{gpbfrontiers86}
  G.P. Beretta, A general nonlinear evolution equation for irreversible conservative approach to stable equilibrium, in \emph{Frontiers of Nonequilibrium Statistical Physics, Proc. NATO ASI, Santa Fe, 1984}, edited by G.T. Moore and M.O. Scully (Plenum Press, New York, 1986), NATO ASI Series B: Physics, Vol. 135, pp. 193-204.




\bibitem{ASME86}
G.P. Beretta, A new approach to constrained-maximization nonequilibrium problems, in \emph{Computer-Aided Engineering of Energy Systems: Second Law Analysis and Modeling}, Edited by R.A. Gaggioli, ASME Book H0341C-AES, Vol. 3, pp. 129-134 (1986).


\bibitem{ASME87Boston}
G.P. Beretta, Dynamics of smooth constrained approach to maximum entropy, in \emph{Second Law Analysis of Thermal Systems}, Edited by M.J. Moran and E. Sciubba, ASME Book I00236, pp. 17-24 (1987).


\bibitem{ASME87Rome}
G.P. Beretta, Steepest-ascent constrained approach to maximum entropy, in \emph{Second Law Analysis of Heat Transfer in Energy Systems}, edited by R.F. Boehm and N. Lior, ASME Book G00390, HTD Vol. 80, pp. 31-38 (1987).

\bibitem{gpB08}
G.P. Beretta, Modeling non-equilibrium dynamics of a discrete probability
  distribution: General rate equation for maximal entropy generation in a
  maximum-entropy landscape with time-dependent constraints, Entropy \textbf{10}, 160--182 (2008).


\bibitem{gpB14}
G.P. Beretta, Steepest entropy ascent model for far-nonequilibrium thermodynamics: Unified implementation of the maximum entropy production principle, Phys. Rev. E \textbf{90}, 042113 (2014).

\bibitem{mG13}
M. Grmela, Extensions of nondissipative continuum mechanics toward complex
  fluids and complex solids, Continuum Mechanics and Thermodynamics \textbf{25}, 55--75 (2013).

\bibitem{mG14}
M. Grmela, Contact Geometry of Mesoscopic Thermodynamics and Dynamics, Entropy \textbf{16}, 1652--1686 (2014).

\bibitem{hcO05}
H.C. \"Ottinger, {\em Beyond Equilibrium Thermodynamics} (Wiley, Hoboken, NJ, 2005).

\bibitem{gpB84}
G.P. Beretta, E.P. Gyftopoulos, J.L. Park, and G.N. Hatsopoulos, Quantum thermodynamics. A new equation of motion for a single constituent of matter, Il Nuovo Cimento B \textbf{82}, 169--191 (1984).

\bibitem{gpB85}
G.P. Beretta, E.P. Gyftopoulos, and J.L. Park, Quantum thermodynamics. A new equation of motion for a general quantum system, Il Nuovo Cimento B \textbf{87}, 77--97 (1985).

\bibitem{jM85}
J. Maddox, Uniting mechanics and statistics, Nature \textbf{316}, 11 (1985).



\bibitem{gpB85b}
G.P. Beretta, Entropy and irreversibility for a single isolated two-level system: new individual quantum states and new nonlinear equation of motion, Int. J. Theor. Phys. \textbf{24}, 119--134 (1985).

\bibitem{gpB85c}
G.P. Beretta, Effect of irreversible atomic relaxation on resonance fluorescence, absorption, and stimulated emission, Int. J. Theor. Phys. \textbf{24}, 1233--1258 (1985).

\bibitem{gpB06}
G.P. Beretta, Nonlinear model dynamics for closed-system, constrained, maximal-entropy-generation relaxation by energy redistribution, Phys. Rev. E \textbf{73}, 026113 (2006).


\bibitem{foot1}
We do not enter here into the details of the nonetheless important question of what is the operative definition of thermodynamics entropy and what specific functional represents it in a given level of description \cite{ZB14,LY13,BZ11}.

\bibitem{ZB14}
E. Zanchini and G.P. Beretta, Recent progress in the definition of thermodynamic entropy, Entropy \textbf{16}, 1547--1570 (2014).

\bibitem{LY13} E.H. Lieb  and J. Yngvason, The entropy concept for nonequilibrium states, Proc. Royal Soc. A \textbf{469}, 20139408 (2013).

\bibitem{BZ11}
G.P. Beretta and E. Zanchini, Rigorous and general definition of thermodynamic entropy, in {\em Thermodynamics}, edited by M. Tadashi (InTech, Rijeka, Croatia, 2011), pp. 23--50. ArXiv: quant-ph/1010.0813.

\bibitem{GB05}
E.P. Gyftopoulos and G.P. Beretta, {\em Thermodynamics: Foundations and Applications} (Dover Mineola, 2005; 1st ed., Macmillan, New York, 1991).

\bibitem{HK65}
G.N. Hatsopoulos and J.H. Keenan, {\em Principles of General Thermodynamic} (Wiley, New York, 1965).

\bibitem{HOT81} F. Hiai, M. Ohya, and M. Tsukada, Sufficiency, KMS Condition and Relative Entropy in Von Neumann Algebras, Pacific J. Math. \textbf{96}, 99--109 (1981).

\bibitem{OR03}
A. Odzijewicz and T.S. Ratiu, Banach Lie-Poisson spaces and reduction, Comm. Math. Phys. \textbf{ 243} 1--54 (2003).


\bibitem{foot2}
	Generalizations of this point have been proposed especially in the context of \emph{contact manifolds} \cite{rH73,mG08}.

\bibitem{mG08}
M. Grmela, Contact geometry of nonequilibrium thermodynamics,  {\em Meeting the Entropy Challenge: An International Thermodynamics   Symposium in Honor and Memory of Professor Joseph H. Keenan}, edited by G.P. Beretta, A.F. Ghoniem, and G.N. Hatsopoulos, AIP Conf. Proc. Series \textbf{1033},  235--240 (2008).


\bibitem{noteSymplectic}
More precisely, symplectic leaves are the  equivalence classes of points that can be matched by a piecewise smooth curve, each element of which is a trajectory of a locally defined Hamiltonian vector field \cite{MR03}.


\bibitem{mG93}
M. Grmela, Thermodynamics of driven systems, Phys. Rev. E  \textbf{48}, 919--930 (1993).

\bibitem{hcO06}
H.C. \"Ottinger, Nonequilibrium thermodynamics for open systems, Phys. Rev. E  \textbf{73}, 036126 (2006).

\bibitem{rM06}
R. Montgomery, {\em A Tour of Subriemannian Geometries, Their Geodesics and
  Applications}, Mathematical Surveys and Monographs (American Mathematical Society, Providence, RI,  2006).


\bibitem{foot3}
A distribution is a subset of the tangent bundle of a manifold satisfying certain simple properties. See \cite{MR03} for the example of Poisson structures in Classical Mechanics.

\bibitem{jL13}
J.M. Lee, {\em Introduction to Smooth Manifolds} (Springer, New York, 2013).

\bibitem{pS74}
P. Stefan, Accessible sets, orbits, and foliations with singularities,  Proc. London Math. Soc. \textbf{s3-29}, 699--713 (1974).

\bibitem{hjS73}
H.J. Sussmann, Orbits of families of vector fields and integrability of
  distributions, Trans Am. Math. Soc. \textbf{  180}, 171--188 (1973).


\bibitem{JacobyFootnote}Indeed, the Jacobi identity \eqref{JacobyPoisson} is a particular case of the Leibniz identity \eqref{Leibniz} for skew-symmetric brackets. \Eq\eqref{Leibniz} is called \qsl Leibniz identity\qsr because it is nothing else than the usual Leibniz product rule \eqref{Leibniz1} for derivations, where here the derivation is \qsl right multiplication by $C$\qsr.

\bibitem{rL86}R.D. Levine,  Geometry in Classical Statistical Thermodynamics, J. Chem. Phys. \textbf{84}, 910--916 (1986).

\bibitem{GKZ04}A.N. Gorban, I.V. Karlin, and A.Y. Zinovyev, Constructive methods of invariant manifolds for kinetic problems, Phys. Rep. \textbf{396}, 197--403 (2004).

\bibitem{Keck12}G.P. Beretta, J.C. Keck, M. Janbozorgi, and H. Metghalchi, The Rate-Controlled Constrained-Equilibrium Approach to Far-From-Local-Equilibrium Thermodynamics, Entropy \textbf{14}, 92--130 (2012).

\bibitem{Caticha01}N. Caticha and E.A. de Oliveira, Gradient descent learning in and out of equilibrium, Phys. Rev. E \textbf{63}, 061905 (2001).

\bibitem{Caticha04}A. Caticha and R. Preuss, Maximum entropy and Bayesian data analysis: Entropic prior distributions, Phys. Rev. E \textbf{70}, 046127 (2004).

\bibitem{Caticha11}A. Caticha, Entropic dynamics, time and quantum theory, J. Phys. A: Math. Theor. \textbf{44}, 225303 (2011).

\bibitem{mG86}
  M. Grmela, Bracket formulation of diffusion-convection equations, Physica D \textbf{21}, 179--212 (1986).

\bibitem{hcO97}
H.C. \"Ottinger, GENERIC formulation of Boltzmann's kinetic equation, J. Non-\Eq Thermod. \textbf{22}, 386--391 (1997).

\bibitem{SS87} S. Sieniutycz, From a Least Action Principle to Mass Action
Law and Extended Affinity, Chem. Eng. Sci. \textbf{42}, 2697--2711 (1987).

\bibitem{GPBNH13}
G.P. Beretta and N.G. Hadjiconstantinou, Steepest entropy ascent models of the boltzmann equation. comparisons
  with hard-sphere dynamics and relaxation-time models for homogeneous
  relaxation from highly non-equilibrium states, in {\em Proceedings of the ASME 2013 International Mechanical
  Engineering Congress and Exposition, San Diego, November 2013}, paper IMECE2013-64905.

\bibitem{cC88}
C. Cercignani, {\em The Boltzmann Equation and Its Applications} (Springer-Verlag, New York, 1988).

\bibitem{cC90}
C. Cercignani, {\em Mathematical Methods in Kinetic Theory} (Plenum Press, New York, 1990).

\bibitem{KHO01}
M. Kr\"oger, M. H\"utter, and H.C. \"Ottinger, Symbolic test of the Jacobi identity for given generalized 'Poisson' bracket, Computer Phys. Comm. \textbf{137}, 325--340 (2001).

\end{thebibliography}
\end{document}